%% file: BMY.tex
\documentstyle{mn}
\input{psfig.sty}
\input{macros_vdbosch}
\begin{document}


\title[Towards Cosmological Concordance on Galactic Scales]
      {Towards Cosmological Concordance on Galactic Scales}
\author[van den Bosch, Mo \& Yang]
       {Frank C. van den Bosch$^{1}$, H.J. Mo$^{1,2}$, and Xiaohu Yang$^{1,3}$
        \thanks{E-mail: vdbosch@mpa-garching.mpg.de}\\
        $^1$Max-Planck-Institut f\"ur Astrophysik, Karl Schwarzschild
         Str. 1, Postfach 1317, 85741 Garching, Germany\\
        $^2$Department of Physics and Astronomy, University of
         Massachusetts, Amherst, MA 01003-4525, USA\\
        $^3$Center for Astrophysics, University of Science and Technology
         of China, Hefei, Anhui 230026, China}


\date{}

\pagerange{\pageref{firstpage}--\pageref{lastpage}}
\pubyear{2000}

\maketitle

\label{firstpage}


\begin{abstract}
  We use  the observed abundance  and clustering of galaxies  from the
  2dF  Galaxy   Redshift  Survey  to  determine   the  matter  density
  $\Omega_m$ and the linear amplitude of mass fluctuations $\sigma_8$.
  We use a method based  on the conditional luminosity function, which
  allows straightforward computation of the luminosity dependent bias,
  $b$,  of galaxies  with respect  to the  matter  distribution.  This
  allows us to break the degeneracy between bias and $\sigma_8$, which
  has hampered previous attempts  of using large scale structure (LSS)
  data to determine $\sigma_8$.   In addition, it allows the inclusion
  of constraints  on the redshift space distortion  parameter $\beta =
  \Omega_m^{0.6}  / b$,  and  yields average  mass-to-light ratios  as
  function of halo  mass.  Using only the luminosity  function and the
  correlation lengths as function  of luminosity we obtain constraints
  on $\Omega_m$ and  $\sigma_8$ that are in good  agreement with COBE. 
  Models with low $\Omega_m$ and high $\sigma_8$ as well as those with
  high $\Omega_m$ and  low $\sigma_8$ are ruled out  because they over
  (under) predict the amount of clustering, respectively.  We find the
  cluster  mass-to-light ratio,  $\langle  M_{\rm vir}/L  \rangle_{\rm
    cl}$,  to  be  strongly  correlated with  $\sigma_8$.   Using  the
  additional  constraints $\langle  M_{\rm vir}/L  \rangle_{\rm  cl} =
  (350  \pm 70)  h \MLsun$  and $\beta  = 0.49  \pm 0.09$  as Gaussian
  priors significantly tightens the constraints and allows us to break
  the  degeneracy   between  $\Omega_m$  and   $\sigma_8$.   For  flat
  $\Lambda$CDM  cosmologies  with  scale-invariant  power  spectra  we
  obtain  that  $\Omega_m  =  0.27^{+0.14}_{-0.10}$  and  $\sigma_8  =
  0.77^{+0.10}_{-0.14}$  (both  95\%  CL).   Adding  constraints  from
  current  CMB   data,  and  extending   the  analysis  to   a  larger
  cosmological   parameter   space,  we   obtain   that  $\Omega_m   =
  0.25^{+0.10}_{-0.07}$ and $\sigma_8 = 0.78 \pm 0.12$ (both 95\% CL).
  Thus,  we  find strong  indications  that  both  the matter  density
  $\Omega_m$ and the mass  variance $\sigma_8$ are significantly lower
  than  their  ``standard'' concordance  values  of  $0.3$ and  $0.9$,
  respectively.  We show that  cosmologies with $\Omega_m \simeq 0.25$
  and  $\sigma_8 \simeq 0.75$,  as favored  here, predict  dark matter
  haloes that  are significantly less centrally  concentrated than for
  the standard $\Lambda$CDM concordance cosmology.  We argue that this
  may solve both the problem with the rotation curves of dwarf and low
  surface   brightness   galaxies,  as   well   as   the  problem   of
  simultaneously  matching  the  galaxy  luminosity function  and  the
  Tully-Fisher zero-point.
\end{abstract}


\begin{keywords}
cosmology: theory ---
cosmology: cosmological parameters ---
galaxies: formation ---
galaxies: halos ---
large-scale structure of universe ---
dark matter.
\end{keywords}


\section{Introduction}
\label{sec:intro}

The most  popular  cosmological models are  variants of  the cold dark
matter   (CDM) paradigm within   which structure  grows from adiabatic
perturbations imprinted  during   an early  inflationary era.    These
density  perturbations  therefore describe the  initial conditions for
structure formation, and determining the corresponding power spectrum,
$P(k)$, is  one of the holy grails  in astrophysics.  In this paper we
assume that $P(k)$   can  be  well  approximated as  a  power-law  (as
predicted  by  simple   inflationary models), and   use   the observed
abundance  and clustering  properties   of galaxies to constrain   its
normalization.  We  only  consider inflationary  CDM cosmologies  with
adiabatic, scalar-only density  perturbations.  In addition, we assume
that neutrinos add  a negligible mass  to the cosmological budget  and
that the vacuum energy is described by a cosmological constant.  These
cosmologies are  characterized   by  6  parameters:  the   mass/energy
densities (in   terms  of the  critical   density)  of  the   baryons,
$\Omega_b$, the cold  dark   matter, $\Omega_c$ and the   cosmological
constant  $\Omega_{\Lambda}$,  the   Hubble parameter   $h =  H_0/(100
\kmsmpc)$,   and  the   spectral   index,  $n_s$, and   normalization,
$\sigma_8$, of the initial power spectrum.  We define $\Omega_m \equiv
\Omega_b +  \Omega_c$ as the matter  density and $\Omega_K \equiv  1 -
\Omega_m - \Omega_{\Lambda}$ as a measure of the spatial curvature.

Recent years have seen a  tremendous improvement in the constraints on
these  cosmological  parameters, with  a  clear concordance  cosmology
emerging.   The  location of  the  first  peak  in the  angular  power
spectrum of cosmic microwave background (CMB) temperature fluctuations
strongly suggests  a flat  Universe, for which  $\Omega_K =  0$ (e.g.,
Balbi  \etal 2000;  Lange \etal  2001; Pryke  \etal  2002; Netterfield
\etal  2002; Ruhl \etal  2002).  An  array of  different observational
data,  most predominantly  that of  high redshift  supernovae  (SN) Ia
(e.g.,  Riess  \etal  1998;  Perlmutter  \etal  1999),  indicate  that
$\Omega_{\Lambda}  \sim 0.7$.   The contribution  from the  baryons is
determined  from Big-Bang nucleosynthesis  models and  measurements of
the  primeval deuterium  abundance to  be  $\Omega_b h^2  = 0.020  \pm
0.001$  (Burles,  Nollett  \&  Turner  2001).   This  is  in  striking
agreement with $\Omega_b h^2 = 0.022 \pm 0.004$ obtained directly from
recent CMB  anisotropy measurements  (Hanany \etal 2000;  de Bernardis
\etal 2002;  Pryke \etal 2002;  Netterfield \etal 2002).  The  HST Key
project has  constrained the  Hubble constant to  $h = 0.72  \pm 0.08$
(Freedman \etal  2001).  Using  this as a  prior, the CMB  data itself
yields  $\Omega_m = 0.3  \pm 0.1$  (e.g., Rubi\~no-Martin  \etal 2002;
Lewis \&  Bridle 2002),  again in excellent  agreement with the  SN Ia
data and  with a  combined analysis of  CMB and large  scale structure
(LSS) data from  the 2dF Galaxy Redshift Survey  (2dFGRS) which yields
$h=0.67  \pm 0.05$  and $\Omega_m  =  0.31 \pm  0.06$ (Percival  \etal
2002).   Finally,  CMB anisotropy  measurements  have constrained  the
spectral   index   of  the   initial   power   spectrum   to  $n_s   =
1.01^{+0.08}_{-0.06}$ (Pryke \etal  2002), in excellent agreement with
the inflation  paradigm which predicts  values of $n_s$ close  to, but
not  necessarily  equal  to,  unity.   Shortly after  this  paper  was
submitted the WMAP-team published their  first-year of data on the CMB
background anisotropies, confirming  and strengthening the CMB results
obtained thus far (Bennett \etal 2003; Spergel \etal 2003).

All these different,  mutually consistent, results seem to  point to a
cosmology with $(\Omega_m,  \Omega_{\Lambda}, \Omega_b, h, n_s) \simeq
(0.3,  0.7,  0.04,   0.7,  1.0)$,  which  has  become   known  as  the
``concordance'' cosmology. Unfortunately, the normalization parameter,
$\sigma_8$, has not yet  been determined with sufficient accuracy that
it can be  included in this list of  ``concordance'' parameters.  Most
attempts to determine $\sigma_8$  use either the observed abundance of
clusters of galaxies,  or the cosmic shear measured  from weak lensing
studies.  Both methods are actually dependent on a combination of both
$\sigma_8$ and  $\Omega_m$ and virtually all  existing constraints are
highly  degenerate in  these  two parameters.   The cluster  abundance
method has the advantage that  it measures the power spectrum directly
at the  scale of interest  (i.e., $\sim 8  h^{-1} \Mpc$) such  that no
extrapolation is required which depends sensitively on the {\it shape}
of the power-spectrum.  The obvious downside of this method is that it
requires  accurate mass  estimates of  individual clusters,  which can
introduce large uncertainties.  The  weak lensing method has the clear
advantage that it  directly probes the distribution of  mass, but with
the  disadvantage   that  cosmic  shear   measurements  are  extremely
difficult. Detailed  overviews and discussions of the  various pro and
cons of  each method, as well  as detailed comparisons  of the results
can be  found in Jarvis  \etal (2003), Smith \etal  (2002a), Pierpaoli
\etal (2002) and references therein.

For the  concordance cosmology ($\Omega_m=0.3$)  current estimates for
$\sigma_8$ range  from $\sim 0.6$  (e.g., Borgani \etal 2001; Reiprich
\& B\"ohringer  2002; Seljak 2002a; Viana,  Nichol \& Liddle  2002) to
$\sim  1.0$ (e.g., Bacon  \etal 2002;  Fan \&  Bahcall 1998; Pen 1998;
Pierpaoli, Scott \&   White  2001;  van   Waerbeke \etal   2002),   an
uncertainty that  is much larger  than the  typical statistical errors
quoted on individual estimates.  Yet,  despite this large uncertainty,
the   vast majority of   numerical simulations and/or galaxy formation
models seem  to adopt $\sigma_8=0.9$  as  the ``standard'' value.  The
reason  is that until fairly  recently most  studies obtained similar,
reasonably consistent results that are well represented by
\begin{equation}
\label{oldclus}
\sigma_8 \Omega_m^{0.5} = 0.5 \pm 0.05
\end{equation}
or $\sigma_8 \simeq 0.9$ for $\Omega_m=0.3$ (Edge \etal 1990; Henry \&
Arnaud 1991;  Bahcall \&  Cen 1992, 1993;  White, Efstathiou  \& Frenk
1993; Kitayama \& Suto 1996; Eke,  Cole \& Frenk 1996; Viana \& Liddle
1996; Pen  1998; Markevitch 1998;  Henry 2000).  Only the  more recent
estimates listed  above have suggested  significantly different values
for $\sigma_8$.

In this  paper we  present a new  method to constrain  $\sigma_8$ (and
$\Omega_m$), based on measurements  of the abundance and clustering of
galaxies. The problem with this kind  of LSS data is that one needs to
transform the  observed distribution of galaxies to  a distribution of
{\it  mass}.   This requires knowledge of the so-called bias parameter
$b$, conveniently defined as 
\begin{equation}
\label{biasdef}
b^2 = {P_{\rm gg}(k) \over P(k)}
\end{equation}
with $P_{\rm  gg}(k)$  the galaxy power spectrum.   Unfortunately, the
bias parameter $b$ depends on scale, galaxy luminosity and galaxy type
(e.g., Kauffmann, Nusser \& Steinmetz 1997; Jing, Mo \& B\"orner 1998;
van den Bosch, Yang \& Mo 2003) and is extremely difficult to measure.
On sufficiently large,  linear  scales, the scale  dependence vanishes
(e.g., Pen 1998) and the {\it shape} of $P_{\rm gg}(k)$ is the same as
that  of  the  matter     power-spectrum.  Therefore,  LSS    data  is
predominantly   used  to constrain  the  shape  parameter $\Omega_m h$
(e.g., Percival \etal 2001;  Efstathiou \etal 2002), while constraints
on the  normalization  $\sigma_8$ require independent  measurements of
the bias $b$.  The method we present here  is based on the conditional
luminosity  function, introduced by  Yang, Mo \&  van den Bosch (2003)
and van den  Bosch, Yang \& Mo (2003),  and takes  this bias (and  its
luminosity dependence) implicitly into account.   Using data from  the
2dFGRS only we derive  constraints  on $\Omega_m$ and $\sigma_8$  that
are in excellent agreement with similar constraints from CMB data.  We
show that additional constraints on the average mass-to-light ratio of
clusters can significantly  improve these constraints  and argue for a
flat $\Lambda$CDM cosmology with   $\Omega_m \sim 0.25$ and  $\sigma_8
\sim 0.75$.  Dark matter haloes  in such a cosmology are significantly
less concentrated  than in  the  standard $\Lambda$CDM cosmology  with
$\Omega_m =  0.3$ and $\sigma_8   \sim 0.9$. In  fact,  we show that a
small reduction of  $\Omega_m$  and/or $\sigma_8$, as  suggested here,
significantly alleviates two important    problems for the   standard,
$\sigma_8=0.9$, concordance cosmology:  the  claim that the   rotation
curves of     dwarf and low  surface brightness     (LSB) galaxies are
inconsistent   with CDM haloes, and   the  failure of galaxy formation
models to simultaneously   match  the galaxy luminosity function   and
Tully-Fisher zero-point.

This  paper  is organized  as  follows.   In Section~\ref{sec:CLF}  we
present constraints  on $\sigma_8$ and $\Omega_m$ from  2dFGRS data on
the clustering properties  of galaxies.  In Section~\ref{sec:joint} we
combine  these data  with  published data  on  the CMB  and perform  a
6-parameter analysis of  flat cosmologies. In Section~\ref{sec:implic}
we discuss  implications of a  small reduction in both  $\Omega_m$ and
$\sigma_8$  with respect to  the standard  values on  the Tully-Fisher
relation and  the rotation  curves of LSB  galaxies. We  summarize our
results in Section~\ref{sec:concl}.
\begin{table*}
\begin{minipage}{\hdsize}
\caption{Cosmological Models.}
\begin{tabular}{ccccccccc}
   \hline
ID & $\Omega_m$ & $\sigma_8$ & $f_{\rm bar}$ & $t_0$ 
 & $\chi^2(\Phi)$ & $\chi^2(\xi)$ & $\langle M_{\rm vir}/L \rangle_{\rm cl}$ & $\beta$ \\
 (1) & (2) & (3) & (4) & (5) & (6) & (7) & (8) & (9) \\
\hline\hline
\lamA & $0.30$ & $0.90$ & $0.14$ & $13.46$ & $63.0$ & $0.91$ & $512$ & $0.55$ \\
\lamB & $0.30$ & $0.65$ & $0.14$ & $13.46$ & $64.6$ & $1.83$ & $265$ & $0.40$ \\
\lamC & $0.25$ & $0.75$ & $0.16$ & $14.16$ & $63.2$ & $1.26$ & $353$ & $0.43$ \\
\lamD & $0.20$ & $0.65$ & $0.20$ & $15.03$ & $63.2$ & $1.29$ & $268$ & $0.34$ \\
\hline
\end{tabular}
\medskip

Parameters  of four  flat   $\Lambda$CDM cosmologies discussed  in the
text.    Column~(1)   lists the ID,    column~(2)  the matter density,
$\Omega_m$, and column~(3)  the  matter power   spectrum normalization
$\sigma_8$. All four models  have $\Omega_{\Lambda} = 1.0 - \Omega_m$,
a  Hubble constant $h=0.7$, a spectral  index $n_s=1.0$,  and a baryon
matter density $\Omega_b h^2  = 0.02$. The corresponding baryonic mass
fraction $f_{\rm bar} = \Omega_b /\Omega_m$ and age of the Universe in
Gyrs,    $t_0$,  are  listed   in  Columns~(4)  and~(5), respectively.
Columns~(6)    and~(7)    list   the     values   of    $\chi^2(\Phi)$
(equation~[\ref{chisqLF}])             and               $\chi^2(\xi)$
(equation~[\ref{chisqr0}]) of the  best fit model for the  conditional
luminosity function, while the corresponding values for $\langle M{\rm
vir}/L     \rangle_{\rm cl}$    (in      $h   \MLsun$)  and    $\beta$
(equation~[\ref{betaobs}])   are   listed   in  columns~(8)   and~(9),
respectively.

\end{minipage}
\end{table*}

For the purpose of facilitating the  discussion that follows, we shall
compare four flat  $\Lambda$CDM cosmologies  with different $\Omega_m$
and $\sigma_8$.  Table~1 lists  a number  of  characteristics of these
models, including  the  baryonic  mass fraction   and the age   of the
Universe.  All four  models have $\Omega_b  h^2 = 0.02$,  $h=0.7$, and
$n_s = 1.0$.  Model \lamA is  the standard $\Lambda$CDM cosmology with
$\Omega_m=0.3$  and   $\sigma_8=0.9$.    Model  \lamB   has   the same
concordance value of  $\Omega_m$ but with a  lower $\sigma_8$.  Models
\lamC and \lamD, finally, have  both $\Omega_m$ and $\sigma_8$ lowered
with respect to the standard values.

\section{Constraints from the abundance and clustering of galaxies}
\label{sec:CLF}

\subsection{The Conditional Luminosity Function}
\label{sec:condlf}

Yang, Mo \&  van den Bosch (2003,  hereafter YMB03) and van den Bosch,
Yang \& Mo (2003; hereafter BYM03) presented a  new method to link the
distribution  of galaxies to that of  dark matter haloes.  This method
is based on modeling of the conditional luminosity function (hereafter
CLF), $\Phi(L \vert  M) {\rm d}L$, which  gives the average  number of
galaxies with luminosity  $L \pm {\rm d}L/2$ that  reside in a halo of
mass $M$.  This CLF is the direct link between  the halo mass function
$n(M)  {\rm d}M$, specifying the  comoving number density of haloes of
mass  $M$, and the   galaxy  luminosity function  $\Phi(L) {\rm  d}L$,
specifying  the comoving number   density of galaxies  with luminosity
$L$, through
\begin{equation}
\label{clf}
\Phi(L) = \int_{0}^{\infty} \Phi(L \vert M) \, n(M) \, {\rm d}M
\end{equation}
Alternatively, one  may link galaxies to their  dark matter  haloes by
specifying the  distribution of halo   masses associated with  a given
galaxy luminosity  (see e.g., Guzik   \& Seljak  2002). In fact,  both
methods are related to each other  via Bayes' theorem (see eq.~[54] in
YMB03).

In CDM  cosmologies, more massive  haloes are more  strongly clustered
(Cole  \& Kaiser  1989;  Mo \&  White  1996, 2002).   This means  that
information  on  the  clustering  strength  of galaxies  (of  a  given
luminosity) contains information about  the characteristic mass of the
halo in which they reside.  Therefore, an observed luminosity function
$\Phi(L)$  combined with measurements  of the  galaxy-galaxy two-point
correlation   function  $\xi_{\rm  gg}(r,L)$   {\it  as   function  of
luminosity} puts stringent constraints on $\Phi(L \vert M)$ (see YMB03
and  BYM03).  In  addition, the  CLF allows  one to  compute  the {\it
average}, total luminosity of galaxies in a halo of mass $M$
\begin{equation}
\label{avlum}
\langle L \rangle(M) = \int_{0}^{\infty} \Phi(L \vert M) \, L \, {\rm d}L
\end{equation}
and   therewith  the   average  mass-to-light   ratios   $\langle  M/L
\rangle(M)$.   These can  be compared  with  independent measurements,
providing  further constraints  on  $\Phi(L \vert  M)$. 

For a given CLF  the  luminosity function $\Phi(L)$ follows   directly
from  equation~(\ref{clf})  while, at sufficiently  large  separations
$r$, the two-point correlation function is given by
\begin{equation}
\label{ggcf}
\xi_{\rm gg}(r,L) = \bar{b}^2(L) \, \xi_{\rm dm}(r)
\end{equation}
Here $\xi_{\rm  dm}(r)$ is the  dark matter mass correlation function,
and $\bar{b}(L)$ is the  average {\it bias}  of galaxies of luminosity
$L$, which derives from the CLF according to
\begin{equation}
\label{biasmod}
\bar{b}(L) = {1 \over \Phi(L)} \int_{0}^{\infty} \Phi(L
\vert M) \, b(M) \, n(M) \, {\rm d}M.
\end{equation}
with $b(M)$ the bias of dark matter haloes  of mass $M$ (see BYM03 for
details).

The mass function $n(M)$ of dark matter haloes at $z=0$ can be written
in the form
\begin{equation}
\label{halomf}
n(M) \, {\rm d}M = {\bar{\rho} \over M^2} \nu f(\nu) \,
\left| {{\rm d} {\rm ln} \sigma \over {\rm d} {\rm ln} M}\right|
{\rm d}M.
\end{equation}
Here $\bar{\rho}$ is the mean matter density of the Universe at $z=0$,
$\nu = \delta_c/\sigma(M)$,  $\delta_c$  is the critical   overdensity
required for collapse at $z=0$, $f(\nu)$ is a  function of $\nu$ to be
specified below, and $\sigma(M)$ is the linear rms mass fluctuation on
mass scale $M$, which is given by the linear power spectrum of density
perturbations $P(k)$ as
\begin{equation}
\label{variance}
\sigma^2(M) = {1 \over 2 \pi^2} \int_{0}^{\infty} P(k) \;
\widehat{W}_{M}^2(k) \; k^2 \; {\rm d}k,
\end{equation}
where $\widehat{W}_{M}(k)$  is the Fourier  transform of the smoothing
filter on mass  scale   $M$\footnote{Throughout, we adopt   a  spatial
top-hat  filter for which  $\widehat{W}_{M}(k)  = 3 (k R)^{-3}  \left[
\sin(k R) - k  R \cos(k R)\right]$ with $M$  and $R$ related according
to $M = 4 \pi \bar{\rho} R^3 / 3$.}.

We  use  the  extended  Press-Schechter theory  with  the  ellipsoidal
collapse corrections of Sheth, Mo \& Tormen (2001) and write
\begin{equation}
\label{fnuST}
\nu \, f(\nu) = 0.644 \,\left(1 + {1\over \nu'{^{0.6}}}\right)\
\left({\nu'{^2}\over 2\pi}\right)^{1/2}
\exp\left(-{\nu'{^2} \over 2}\right)\,
\end{equation}
and
\begin{eqnarray}
\label{bm}
b(M) & = & 1 + {1\over\sqrt{a}\delta_{c}(z)}
\Bigl[ \sqrt{a}\,(a\nu^2) + \sqrt{a}\,b\,(a\nu^2)^{1-c} - \nonumber \\
& & {(a\nu^2)^c\over (a\nu^2)^c + b\,(1-c)(1-c/2)}\Bigr],
\end{eqnarray}
with $\nu'=0.841\,\nu$, $a=0.707$, $b=0.5$ and $c=0.6$.  The resulting
mass function and correlation function of dark matter haloes have been
shown to be in excellent agreement with numerical simulations, as long
as  halo masses  are defined  as the  masses inside  a sphere  with an
average overdensity of  about $180$ (Jing 1998; Sheth  \& Tormen 1999;
Jenkins  \etal  2001; White  2002).   Therefore,  in  what follows  we
consistently  use  that definition  of  halo  mass  when referring  to
$M$.  In Section~\ref{sec:mlcl} we  also define  the virial  mass, for
which we use  the symbol $M_{\rm vir}$. Finally, we  use the CDM power
spectrum  $P(k)$ of  Efstathiou, Bond  \&  White (1992)  with a  shape
parameter
\begin{equation}
\label{gamma}
\Gamma = \Omega_m \, h \, {\rm exp}(-\Omega_b - \sqrt{2h}
\Omega_b/\Omega_m)
\end{equation}
(Sugiyama 1995), and compute $\xi_{\rm dm}(r)$ from
\begin{equation}
\label{dmcorrfunc}
\xi_{\rm dm}(r) = \int_{0}^{\infty} \Delta^2_{\rm nl}(k) \, {\sin(k r)
\over k r} \, {{\rm d}k \over k}
\end{equation}
with $\Delta_{\rm nl}(k)$ the evolved non-linear power spectrum of the
dark matter mass distribution, for which we use the fitting formula of
Smith \etal (2002b).
\begin{figure*}
\centerline{\psfig{figure=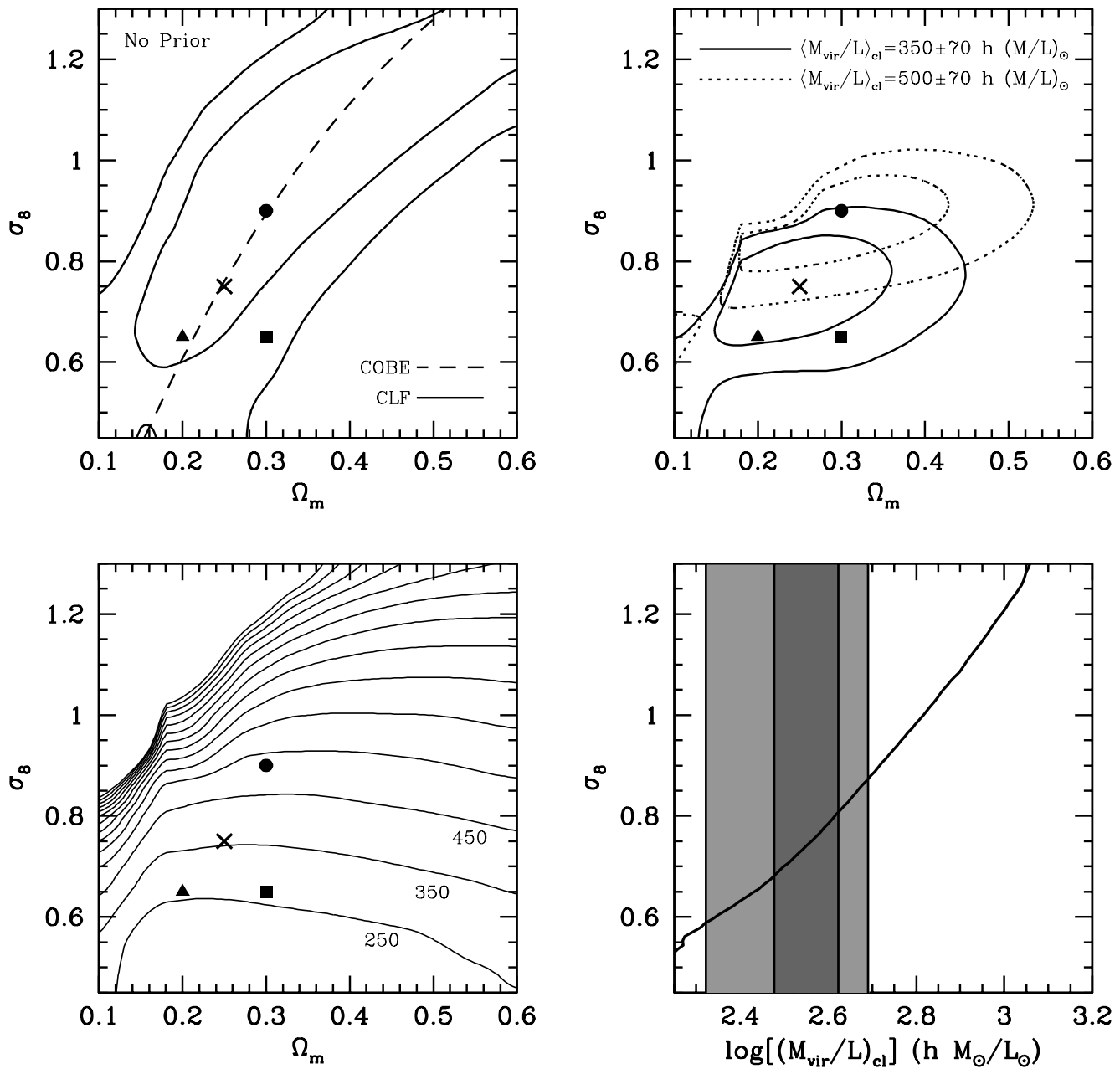,width=\hdsize}}
\caption{The upper left  panel plots the 68 and  95 percent confidence
levels of ${\cal  L}_{\rm CLF}(\Omega_m,\sigma_8)$ (solid lines).  For
comparison,  the dashed  line  indicates the  value  of $\sigma_8$  as
function of $\Omega_m$ obtained from the COBE normalization for a flat
$\Lambda$CDM  cosmology with  $\Omega_b h^2=0.02$,  $h=0.7$, $n_s=1.0$
and $\tau=0.0$. The various symbols indicate the four models discussed
in the  text: \lamA (solid  dot), \lamB (solid square),  \lamC (cross)
and \lamD (solid triangle). The solid and dotted contours in the upper
right panel  show the  68 and 95  percent confidence levels  of ${\cal
L}_{\rm CLF}(\Omega_m,\sigma_8)$  when combined with  a Gaussian prior
on the mass-to-light ratio of  clusters (as indicated). The lower left
panel plots  contours of constant $\langle  M_{\rm vir}/L \rangle_{\rm
cl}$ at values of $250, 350, 450,...,1550 h \MLsun$.  For clarity, the
first  three contours are  labelled.  Finally,  the lower  right panel
plots  the   value  of   $\sigma_8$  that  maximizes   ${\cal  L}_{\rm
CLF}(\Omega_m,\sigma_8)$  for  given  $\Omega_m$  as function  of  the
$\langle  M_{\rm   vir}/L  \rangle_{\rm  cl}$   of  the  corresponding
model. Clearly,  higher values of $\langle  M_{\rm vir}/L \rangle_{\rm
cl}$  imply  higher  normalizations  $\sigma_8$, as  parameterized  by
equation~(\ref{s8ml}).  The light (dark)  gray regions indicate the 95
(68) percent confidence levels on  the observed $\langle M_{\rm vir}/L
\rangle_{\rm cl}$ (see Section~\ref{sec:mlcl}).}
\label{fig:clf}
\end{figure*}

\subsection{Cosmological Constraints}
\label{sec:constraints}

Both $n(M)$ and $b(M)$ are cosmology dependent.  Whether  or not a CLF
can be found  that results in good fits  to the observed $\Phi(L)$ and
$\xi_{\rm gg}(r,L)$   therefore  depends  on  the   cosmological model
adopted, and this method may therefore be used to place constraints on
cosmological parameters.   In YMB03 we  used  data from  the 2dFGRS to
constrain  the  matter   density  $\Omega_m$ in   a  flat $\Lambda$CDM
cosmology.  We only considered cosmological models consistent with the
recent weak lensing constraints  of Hoekstra, Yee \&  Gladders (2002),
i.e.,   $\sigma_8 \Omega_m^{0.52}   = 0.46$, and    adopted an average
cluster mass-to-light ratio  of $\langle M/L \rangle_{\rm  cl} = 500 h
\MLsun$.  It was shown that  under these conditions the best-fit model
has   $\Omega_0 =  0.3$  (and  thus   $\sigma_8 = 0.9$),  in excellent
agreement with the standard concordance cosmology.

Here  we present  a more  detailed investigation.  Unlike in  YMB03 we
consider $\sigma_8$ and $\Omega_m$ as independent model parameters and
let  $\langle   M/L  \rangle_{\rm  cl}$  be  a   free  parameter  (see
Appendix~A).  We start by considering only flat cosmologies ($\Omega_K
=  0.0$) with  $\Omega_b h^2  = 0.02$,  $h=0.7$ and  $n_s =  1.0$, and
investigate  how  the   observed  $\Phi(L)$  and  $\xi_{\rm  gg}(r,L)$
constrain $\Omega_m$ and $\sigma_8$.  In Section~\ref{sec:joint} below
we  relax  these assumptions  and  perform  a  more detailed  analysis
allowing $\Omega_b  h^2$, $h$,  and $n_s$ to  vary as well.   For each
($\Omega_m$,$\sigma_8$) we determine the CLF that best fits the 2dFGRS
luminosity  function of  Madgwick \etal  (2002) and  the  amplitude of
$\xi_{\rm gg}(r,L)$ at the  correlation lengths $r_0(L)$ obtained from
the 2dFGRS  by Norberg \etal  (2002). Note that throughout  this paper
all luminosities  are in  the photometric $b_J$-band  (uncorrected for
intrinsic absorption  by dust), unless  specifically stated otherwise.
We parameterize $\Phi(L \vert M)$ using a model with 7 free parameters
(see  Appendix~A)  and use  Powell's  multi-dimensional direction  set
method (e.g., Press \etal 1992) to find those parameters that minimize
\begin{equation}
\label{chisq}
\chi^2 = \chi^2(\Phi) + \chi^2(\xi)
\end{equation}
and  thus maximize  the  likelihood ${\cal  L}_{\rm  CLF} \equiv  {\rm
e}^{-\chi^2/2}$. Here the first term
\begin{equation}
\label{chisqLF}
\chi^2(\Phi) = \sum_{i=1}^{N_{\Phi}}
\left[ {\Phi(L_i) - \hat{\Phi}(L_i) \over \Delta \hat{\Phi}(L_i)} \right]^2,
\end{equation}
measures the goodness-of-fit to the observed LF $\hat{\Phi}(L_i)$ with
errors $\Delta \hat{\Phi}(L_i)$.  The second term is defined by
\begin{equation}
\label{chisqr0}
\chi^2(\xi) = \sum_{i=1}^{N_{\xi}}
\left[ {\xi_{\rm gg}(\hat{r}_0(L_i),L_i) - \hat{\xi}_{\rm
gg}(\hat{r}_0(L_i),L_i) \over \Delta \hat{\xi}_{\rm
gg}(\hat{r}_0(L_i),L_i)} \right]^2,
\end{equation}
and   measures  the   goodness-of-fit   to  the   amplitudes  of   the
galaxy-galaxy  correlation functions. Note  that we  compute $\xi_{\rm
gg}(r,L)$  at the {\it  observed} correlation  lengths $\hat{r}_0(L)$.
These   are   compared   to   the  observed   values   $\hat{\xi}_{\rm
gg}(\hat{r}_0(L_i),L_i)$, which by definition are equal to unity.  The
errors  $\Delta \hat{\xi}_{\rm  gg}(\hat{r}_0(L_i),L_i)$  are computed
from  the  errors  on  $\hat{r}_0(L_i)$  and the  power-law  slope  of
$\hat{\xi}_{\rm gg}(r,L_i)$ quoted by Norberg \etal (2002).

The upper left  panel  of Figure~\ref{fig:clf}  plots  the  68  and 95
percent  confidence    levels,   obtained by   integrating   under the
likelihood surface using the flat priors  $0.1 \leq \Omega_m \leq 0.6$
and  $0.45 \leq  \sigma_8 \leq 1.30$.    As is apparent,  the observed
abundance  and     clustering  properties constrain    $\Omega_m$  and
$\sigma_8$ to a fairly broad valley in $(\Omega_m,\sigma_8)$ parameter
space, and all four cosmologies listed  in Table~1 are consistent with
the observations at  better than 95  percent confidence.   Cosmologies
with  small  $\Omega_m$ and large   $\sigma_8$   and those with  large
$\Omega_m$ and small $\sigma_8$ are ruled out at  more than 95 percent
confidence.   For comparison, the  dashed  line indicates the relation
between  $\sigma_8$   and    $\Omega_m$   that  best-fits    the  COBE
data\footnote{For    flat  $\Lambda$CDM  cosmologies   with  $\Omega_b
h^2=0.02$, $h=0.7$, $n_s=1.0$  and zero optical depth to reionization,
obtained using the parameterization of Liddle \etal (1996) and Bunn \&
White (1997)  and the  matter transfer function  of  Eisenstein \&  Hu
(1998).} (Bennett \etal  1996). This COBE constraint  nicely coincides
with the  valley  floor  of ${\cal  L}_{\rm  CLF}(\Omega_m,\sigma_8)$,
therewith  indicating  that both the CMB   anisotropies and the galaxy
clustering indicate the same shape of the power spectrum $P(k)$.

\begin{figure*}
\centerline{\psfig{figure=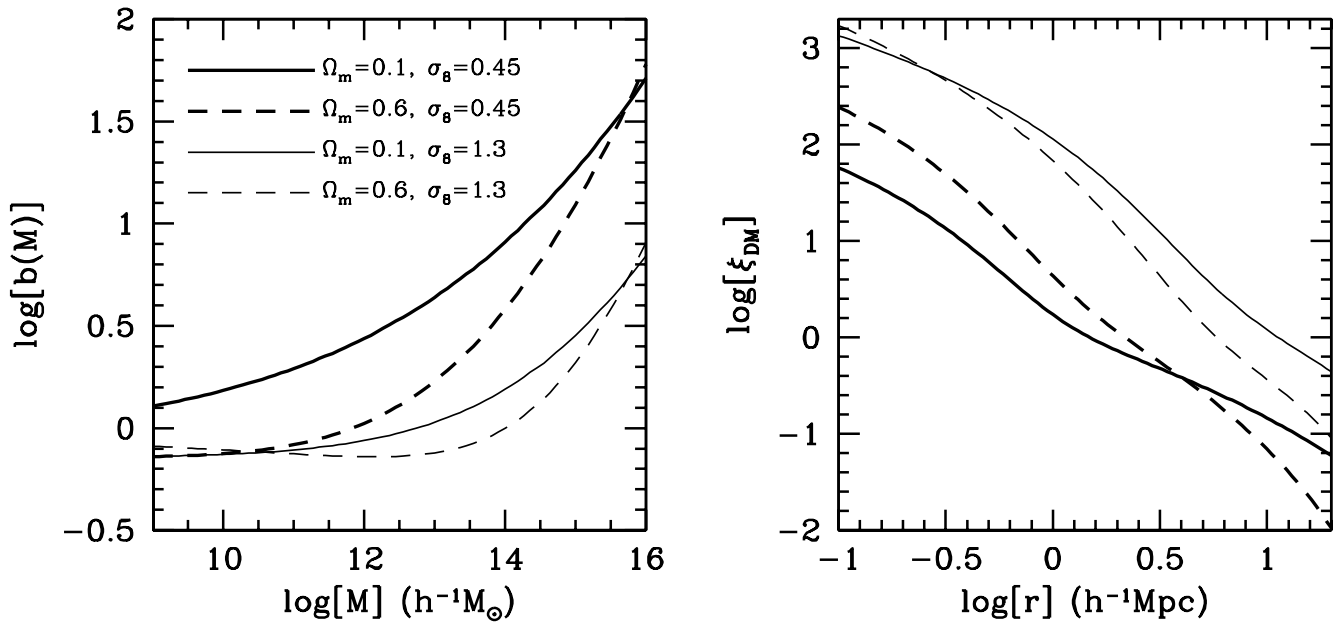,width=\hdsize}}
\caption{The halo bias $b(M)$ (left panel) and dark matter correlation
function $\xi_{\rm DM}(r)$ (right  panel) for four extreme cosmologies
as  indicated.  Note  how   more  massive  haloes  are  more  strongly
clustered,   and   that  $b(M)$   {\it   decreases}  with   increasing
$\sigma_8$.  This   largely  counter-balances  the   increase  in  the
clustering strength  with increasing $\sigma_8$, thus  explaining to a
large extent  why the observed  clustering of galaxies only  puts very
mild  constraints  on  $\sigma_8$   (see  text  for  a  more  detailed
discussion).}
\label{fig:cospar}
\end{figure*}

Understanding   how  $\Phi(L)$   and  $\xi_{\rm   gg}(r,L)$  constrain
$\sigma_8$  and  $\Omega_m$ is  somewhat  complicated.   To guide  the
discussion,  Figure~\ref{fig:cospar} plots the  bias, $b(M)$,  and the
evolved, non-linear correlation function  of the dark matter $\xi_{\rm
DM}(r)$ (equation~[\ref{dmcorrfunc}]) for the four extreme cosmologies
considered.   First of  all, the  fact  that we  demand a  fit to  the
observed  LF hardly  imposes any  constraints at  all;  $\Phi(L)$ only
depends  on the  halo  mass function,  and  one can  always choose  an
appropriate $\Phi(L \vert M)$ so  that one fits $\Phi(L)$ perfectly no
matter what the shape or normalization of $n(M)$.  We only include the
observed LF as  a constraint since it sets  the {\it normalization} of
the CLF, which  allows us to compute mass-to-light  ratios. As we show
below,  this proves  to be  an  important asset.   The constraints  on
$\Omega_m$ and $\sigma_8$ are almost  solely due to the constraints on
$\xi_{\rm gg}(r,L)$.   Typically, increasing $\sigma_8$  increases the
amount   of   clustering.    This   is   immediately   apparent   from
Figure~\ref{fig:cospar} which  shows that $\xi_{\rm  DM}(r)$ increases
drastically from  $\sigma_8=0.45$ to $\sigma_8 =  1.3$.  Therefore, in
order to  keep $\xi_{\rm  gg}(r,L)$ fixed at  the observed  values the
bias $\bar{b}(L)$ needs to be  lowered.  This in turn requires a lower
halo bias  $b(M)$.  However, as is  apparent from equation~(\ref{bm}),
$b(M)$ can not become arbitrary small, thus imposing an upper limit on
$\sigma_8$ for  given $\Omega_m$. To emphasize the  robustness of this
result, consider  the $\Omega_m=0.1$  and $\sigma_8 =  1.3$ cosmology.
At $r  = 5  h^{-1} \Mpc$, roughly  the observed correlation  length of
galaxies with  $-18 \gta M_{b_J}  - 5 {\rm  log} h \gta  -20$ (Norberg
\etal 2002),  the {\it dark  matter} correlation function is  equal to
$4.4$. Using  that $\xi_gg(r = 5  h^{-1} \Mpc) \simeq  1$ implies that
these  galaxies require an  average bias  of $\bar{b}(L)  \simeq 0.48$
(cf.   equation~[\ref{ggcf}]).   This, however,  is  smaller than  the
minimum of $b(M)$ (see Figure~\ref{fig:cospar}).  Therefore, no matter
in what haloes these  galaxies reside, their correlation function will
always be larger than observed. This cosmology is therefore ruled out,
a result that is robust against whatever we assume regarding the CLF.

To  understand  why cosmologies  with  both  high  $\Omega_m$ and  low
$\sigma_8$  are  ruled inconsistent  is  less  trivial. Consider,  for
example,  the cosmology with  $\Omega_m=0.6$ and  $\sigma_8=0.45$.  To
explain  the  galaxy-galaxy correlation  lengths  obtained by  Norberg
\etal (2002), which range from $4.4 h^{-1} \Mpc$ to $8.5 h^{-1} \Mpc$,
requires   values   for   $\bar{b}(L)$   in   the   range   $1.7$   to
$3.2$\footnote{This is easily  verified from equation~(\ref{ggcf}) and
the  dark matter  correlation function  shown  in the  right panel  of
Figure~\ref{fig:cospar}.}.   As is  apparent  from equation~\ref{ggcf}
and   the  left   panel  of   Figure~\ref{fig:cospar},  this   can  be
accomplished by,  for example,  distributing all galaxies  over haloes
within the relatively  narrow range $10^{13} h^{-1} \Msun  \lta M \lta
10^{14} h^{-1} \Msun$.  Therefore, in principle, one should be able to
find  a halo  occupation model  for this  cosmology that  is perfectly
consistent with the  data. The fact that our  model can not accurately
match  the  data must  therefore  reflect  a  restriction due  to  the
parameterization of the CLF. Although our parameterization has partial
observational support,  results in halo occupation  statistics in good
agreement  with semi-analytical  models, and  is robust  against small
changes (see Appendix~B for  a detailed discussion), we emphasize that
at least  part of  the constraints  shown in the  upper left  panel of
Figure~\ref{fig:clf} are due to the particular parameterization of the
conditional  luminosity function  used. Fortunately,  as we  shall see
below,  the lower-left  corner of  the $(\Omega_m,\sigma_8)$-parameter
space is anyway ruled out  by CMB data, and this restriction therefore
does not impact on our results.

\subsection{The mass-to-light ratios of clusters}
\label{sec:mlcl}

One of the 7 free parameters of our $\Phi(L \vert M)$ parameterization
is $\langle M/L \rangle_{\rm cl}$, the  average mass-to-light ratio of
clusters of galaxies    with  $M \geq   10^{14}  h^{-1}  \Msun$   (see
Appendix~A). Note that the halo mass $M$ is defined as the mass within
the radius inside of which the average overdensity  is $180$. In order
to  facilitate a  more   direct comparison with  mass-to-light  ratios
available in the literature we convert $M/L$ to $M_{\rm vir}/L$, where
$M_{\rm vir}$ is defined as the mass within  the virial radius, inside
of which the average density is  $\Delta_{\rm vir}$ times the critical
density. We adopt
\begin{equation}
\label{dvir}
\Delta_{\rm vir} = 18 \pi^2 + 82(\Omega_m-1) - 39(\Omega_m-1)^2
\end{equation}
(Bryan \& Norman 1998). To convert $M$ to $M_{\rm vir}$ we also need a
model for the  halo  concentration  as function    of halo  mass   and
cosmology,  for which  we use the  model of  Eke, Navarro \& Steinmetz
(2001). In what follows, whenever masses correspond to the virial mass
we explicitely write $M_{\rm vir}$.

The  lower left panel  of Figure~\ref{fig:clf}  plots contours  of the
best-fit   value  of   $\langle  M_{\rm   vir}/L   \rangle_{\rm  cl}$.
Typically,  models with  higher  $\sigma_8$ require  higher values  of
$\langle M_{\rm vir}/L \rangle_{\rm cl}$  in order to fit the observed
abundances  and clustering properties  of galaxies.   This is  easy to
understand; consider once  again the $\Omega_m=0.1$ and $\sigma_8=1.3$
cosmology.  As  discussed in the  previous section, this  cosmology is
ruled out  because the dark  matter is too strongly  correlated; there
are simply no  haloes with sufficiently small bias,  $b(M)$, such that
the  galaxy-galaxy correlation  function can  be made  consistent with
observations.   The discrepancy with  the data  is minimized  when all
galaxies reside  in haloes with  $M \simeq 10^{10} h^{-1}  \Msun$, for
which $b(M)$ is mimimal. Clearly,  galaxies need to avoid more massive
haloes  since  otherwise  they  would  overpredict  the  galaxy-galaxy
correlation strength  even more.   This `avoidance' of  massive haloes
explains the extremely large  mass-to-light ratio of clusters for this
cosmology  (see   lower  left  panel   of  Figure~\ref{fig:clf}).   If
$\sigma_8$  is   lowered,  the  dark  matter   becomes  less  strongly
clustered.  Slowly,  galaxies are required  to populate more  and more
massive   haloes  in   order   to  match   the  observed   correlation
lengths. This  implies that  $\langle M_{\rm vir}/L  \rangle_{\rm cl}$
has to  decrease. In addition, lowering  $\sigma_8$ strongly decreases
the  {\it  abundance}  of  massive  haloes.  Since  the  abundance  of
galaxies is  fixed by the observed  LF, one expects,  on average, more
galaxies  per halo.   This again  will  cause $M/L$  to decrease  with
decreasing  $\sigma_8$.  This example  shows  how  constraints on  the
abundance  and  clustering properties  of  galaxies  yields a  cluster
mass-to-light  ratio  that depends  strongly  on  both $\Omega_m$  and
$\sigma_8$.

It is  apparent from Figure~\ref{fig:clf} that  $\langle M_{\rm vir}/L
\rangle_{\rm  cl}$ varies  strongly  with position  along contours  of
constant   ${\cal  L}_{\rm   CLF}$.   This   indicates  that   we  can
significantly tighten the constraints  on $\Omega_m$ and $\sigma_8$ by
using  additional constraints on  $\langle M_{\rm  vir}/L \rangle_{\rm
  cl}$.   Using a  variety of  techniques to  measure  cluster masses,
Bahcall \etal (2000) obtain  $\langle M_{\rm vir}/L_B \rangle_{\rm cl}
=  (330  \pm  77)  h  \MLsun$  (where  $L_B$  is  the  total  $B$-band
luminosity,  uncorrected  for  internal extinction).   Carlberg  \etal
(1996), using the CNOC Cluster Survey, obtain $\langle M_{\rm vir}/L_B
\rangle_{\rm cl} = (363 \pm 65) h \MLsun$ (where we have converted the
Gunn $r$-band luminosity to $L_B$  using $L_r = 1.23 L_B$, see Bahcall
\& Comerford 2002).  Taking the average from these two measurements we
obtain  $\langle M_{\rm  vir}/L_B \rangle_{\rm  cl} =  (350 \pm  70) h
\MLsun$.  Adding this  as a (Gaussian) prior to  the likelihood ${\cal
  L}_{\rm  CLF}(\Omega_m,\sigma_8)$  yields  the  68  and  95  percent
confidence   levels    shown   in    the   upper   right    panel   of
Figure~\ref{fig:clf}  (solid  lines).    As  expected,  the  prior  on
$\langle  M_{\rm  vir}/L \rangle_{\rm  cl}$  drastically tightens  the
constraints   on  $\Omega_m$   and  $\sigma_8$.    Marginalizing  over
$\sigma_8$ and $\Omega_m$  we obtain $\Omega_m = 0.23^{+0.17}_{-0.09}$
and $\sigma_8  = 0.75^{+0.14}_{-0.18}$  (both 95\% CL),  respectively. 
Whereas the \lamB, \lamC and \lamD cosmologies are all consistent with
these constraints  at better than 68 percent  confidence, the standard
\lamA cosmology  is ruled out at  the 99.99 percent  confidence level! 
In fact,  the latter requires  a cluster mass-to-light ratio  of $\sim
512  h \MLsun$  (see  Table~1), more  than  $2 \sigma$  away from  the
observed value.

In order to quantify  the dependence of  $\sigma_8$ on $\langle M_{\rm
vir}/L \rangle_{\rm cl}$  we proceed as  follows.  From ${\cal L}_{\rm
CLF}(\Omega_m,\sigma_8)$   shown    in  the   upper   left   panel  of
Figure~\ref{fig:clf},  we determine  the  value  of   $\Omega_m$  that
maximizes the likelihood for given $\sigma_8$.   The lower right panel
of Figure~\ref{fig:clf} plots $\langle M_{\rm vir}/L \rangle_{\rm cl}$
of  this best-fit model as  function of $\sigma_8$ (i.e., the relation
between $\sigma_8$ and $\langle M_{\rm  vir}/L \rangle_{\rm cl}$ along
the valley floor), which is well approximated by
\begin{equation}
\label{s8ml}
\sigma_8 = 0.73 \left( { \langle  M_{\rm vir}/L \rangle_{\rm cl} \over
350 \, h \, \MLsun } \right)^{0.5}
\end{equation}
The light (dark)   gray areas  correspond    to the  95 (68)   percent
confidence levels on $\langle  M_{\rm vir}/L \rangle_{\rm cl}$.   Note
that, because contours of constant $\langle M_{\rm vir}/L \rangle_{\rm
cl}$  are almost independent of $\Omega_m$  (except  at $\Omega_m \lta
0.20$),  equation~(\ref{s8ml})   is also very    similar  to the  {\it
maximum} $\sigma_8$  that is consistent with  a given  $\langle M_{\rm
vir}/L \rangle_{\rm  cl}$.   Requiring  that  $\langle M_{\rm   vir}/L
\rangle_{\rm  cl}  < 500 h \MLsun$,  for  example, translates  into an
upper limit of $\sigma_8 < 0.87$.

Our  modeling of  the  conditional  luminosity function  combined with
constraints  on the  mass-to-light ratio  of clusters  has resulted in
fairly   stringent constraints on  $\Omega_m$.    It is interesting to
compare  this with an alternative method  to constrain $\Omega_m$ from
$\langle M_{\rm vir}/L \rangle_{\rm cl}$ {\it directly}. One can write
\begin{equation}
\label{omcl}
\Omega_m = {\langle M_{\rm vir}/L \rangle_{\rm cl} \over
\rho_{\rm crit} / \bar{\rho}_L} \, \bar{\cal B}_{\rm cl}
\end{equation}
with $\bar{\rho}_L$ the average   luminosity density in the  Universe,
$\rho_{\rm crit} = 3 H_0^2 / 8  \pi G =  2.78 \times 10^{11} h^2 \Msun
\Mpc^{-3}$ the   critical  density, and  $\bar{\cal  B}_{\rm  cl}$ the
mass-to-light ratio bias  of clusters  of  galaxies.  Note  that  this
`bias'  is  not the  same  as $b$ defined in equation~(\ref{biasdef}).
Using $\langle M_{\rm vir}/L \rangle_{\rm cl} = (350 \pm 70) h \MLsun$
and adopting   $\rho_L   = (1.82  \pm  0.17)  \times  10^{8}  h  \Lsun
\Mpc^{-3}$ (Norberg \etal 2002), we obtain
\begin{equation}
\label{bestfitom}
\Omega_m = (0.23 \pm 0.09) \, \bar{\cal B}_{\rm  cl} 
\end{equation}
which  is  equal  to  the  estimate of  $\Omega_m$  derived  above  if
$\bar{\cal  B}_{\rm cl}  = 1.0$.  Remarkably enough  this is  also the
value  for $\bar{\cal  B}_{\rm cl}$  suggested by  observations, which
indicate that  $M/L$ increases as  a function of  scale up to  $\sim 1
h^{-1} \Mpc$, but then flattens out and remains approximately constant
(Bahcall, Lubin \& Dorman 1995; Bahcall \etal 2000).

\subsection{Peculiar velocities} 
\label{sec:pecvel}

The galaxy-galaxy correlation function $\xi_{\rm gg}(r)$ used above is
derived  from  the   two-dimensional  two-point  correlation  function
$\xi(\sigma,\pi)$, which  measures the excess  probability over random
of finding  a pair of galaxies with  a separation in the  plane of the
sky  $\sigma$  and a  line-of-sight  separation  $\pi$.  Two  peculiar
velocity  effects  distort   this  {\it  redshift}  space  correlation
function with respect to  the real space correlation function $\xi(r)$
and cause  $\xi(\sigma,\pi)$ to be  anisotropic.  On small  scales the
virialized  motion of  galaxies within  dark matter  haloes  cause the
correlation function  to be  extended in the  $\pi$ direction  (the so
called   ``finger-of-God''    effect),   while   on    larger   scales
$\xi(\sigma,\pi)$ is flattened in  the $\pi$ direction due to coherent
infall  of  galaxies  onto  mass concentrations  (Kaiser  1987).   The
amplitude  of  this  Kaiser  effect  depends  on  the  matter  density
$\Omega_m$ and on the biasing of the galaxy distribution. Although the
galaxy  bias  is  a  complicated  function  of  radial  scale,  galaxy
luminosity,  and even galaxy  type (Kauffmann  \etal 1997;  Jing \etal
1998;  YMB03;   BYM03),  on  sufficiently  large   scales  the  radial
dependence  should vanish  (i.e.,  the  bias $b$  is  linear) and  the
flattening of $\xi(\sigma,\pi)$ depends on the parameter
\begin{equation}
\label{betaobs}
\beta = \Omega_m^{0.6}/b.
\end{equation}
Using  data from  the 2dFGRS,  Hawkins  \etal  (2002) recently  used a
multi-parameter fit  to $\xi(\sigma,\pi)$ to obtain  $\beta = 0.49 \pm
0.09$ (68\% CL), the most accurate measurement to date (cf., Hamilton,
Tegmark \&  Padmanabhan  2000;  Taylor  \etal 2001; Outram,   Hoyle \&
Shanks 2001).  As outlined in Hawkins \etal  (2002), the mean redshift
and luminosity  of the  galaxies on which   this measure is  based are
$\bar{z}    =  0.15$ and  $1.4$   times  the characteristic luminosity
$L^{*}$, respectively.
\begin{figure*}
\centerline{\psfig{figure=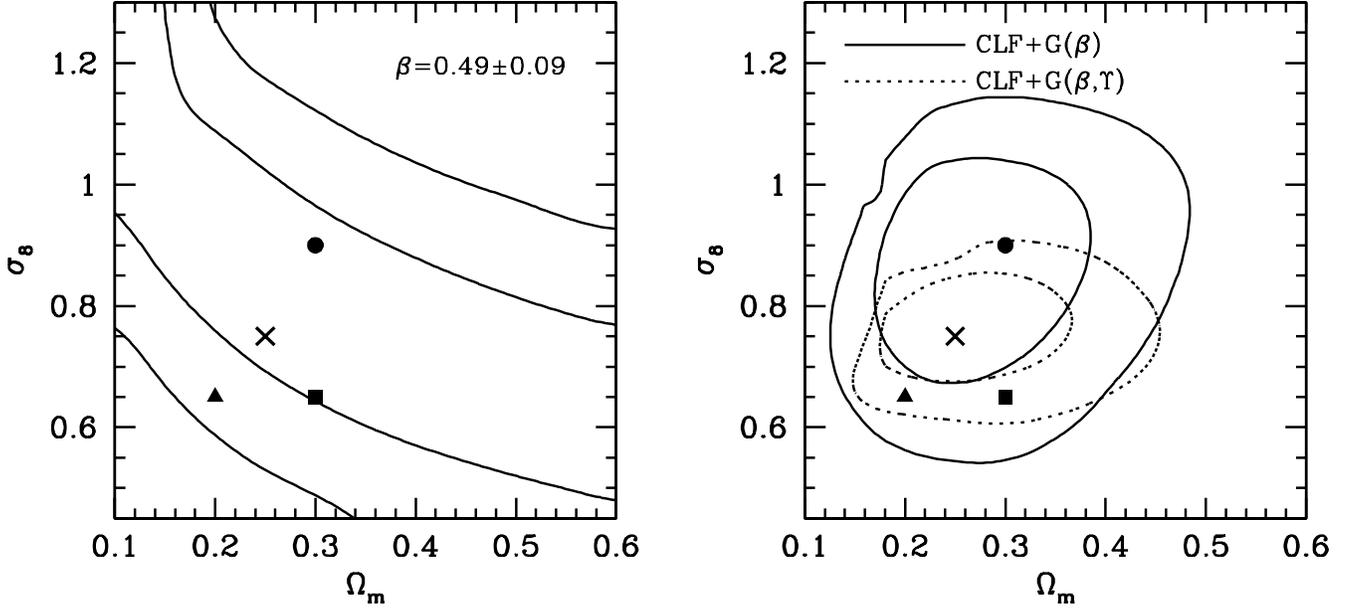,width=\hdsize}}
\caption{The left  panel plots contours  of  constant $\beta$ computed
from  the CLF that   maximizes   ${\cal L}_{\rm CLF}$  for  the  given
$\Omega_m$ and $\sigma_8$. The contours  shown correspond to (from top
to bottom)  $\beta = 0.67, 0.58,  0.40, 0.31$ which are  the 68 and 95
percent confidence  levels of  $\beta$  as obtained by   Hawkins \etal
(2002) from the 2dFGRS.  The right  panel plots the  68 and 95 percent
confidence levels of  ${\cal L}_{\rm CLF}(\Omega_m,\sigma_8)$ obtained
using the priors as indicated.  Here $G(\beta)$ refers to the Gaussian
prior $\beta = 0.49 \pm 0.09$, while $G(\beta,\Upsilon)$ refers to the
combined prior of $G(\beta)$  plus $\langle M_{\rm vir}/L \rangle_{\rm
cl} = (350 \pm 70) h \MLsun$.}
\label{fig:beta}
\end{figure*}

The  problem with  interpreting  constraints on  $\beta$  in terms  of
constraints  on  the  matter  density  $\Omega_m$  is  that  the  bias
parameter  $b$ is  notoriously hard  to measure.   However, it  can be
computed straightforwardly from  the CLF, and as a  function of galaxy
luminosity.  For  each of the  ($\Omega_m,\sigma_8$)-models we compute
$\bar{b}(1.4     L^{*})$    from     the     best-fit    CLF     using
equation~(\ref{biasmod}).   We  convert this  bias  and $\Omega_m$  to
their  corresponding  values at  $\bar{z}  =  0.15$  using the  method
outlined in  BYM03 and compute the likelihood  ${\cal L}_{\beta}$ that
the value of $\beta$ found by Hawkins \etal (2002) originates from the
given   cosmology.     The   contours    in   the   left    panel   of
Figure~\ref{fig:beta}  indicate the  corresponding 68  and  95 percent
confidence levels.  As is  apparent, constraints on $\beta$ only yield
highly  degenerate constraints  on $\Omega_m$  and  $\sigma_8$.  Note,
however,  that  contours of  constant  ${\cal  L}_{\beta}$ run  almost
perpendicular to  those of constant  ${\cal L}_{\rm CLF}$  (cf., upper
left panel of Figure~\ref{fig:clf}).   This indicates that $\beta$ and
$\xi_{\rm  gg}(r,L)$, although  both obtained  from $\xi(\sigma,\pi)$,
constrain $\Omega_m$ and $\sigma_8$ in different ways.  Using $\beta =
0.49   \pm   0.09$   as   a   Gaussian   prior   on   ${\cal   L}_{\rm
CLF}(\Omega_m,\sigma_8)$, results in the  68 and 95 percent confidence
levels  shown  in  the  right panel  of  Figure~\ref{fig:beta}  (solid
contours).  Both  $\Omega_m$ and  $\sigma_8$ are now  well constraint;
marginalizing  over $\sigma_8$  and $\Omega_m$  we obtain  $\Omega_m =
0.27^{+0.16}_{-0.12}$ and $\sigma_8 = 0.87^{+0.22}_{-0.26}$ (both 95\%
CL),  respectively.  Note  that  these constraints  on $\Omega_m$  and
$\sigma_8$  are obtained  from the  2dFGRS alone,  independent  of any
other data set.

Hawkins  \etal (2002)  combined  their constraint  on  $\beta$ with  a
determination  of the  bias obtained  by  Verde \etal  (2002) from  an
analysis of the 2dFGRS bispectrum.   This yielded $\Omega_m = 0.23 \pm
0.09$, consistent  with the results  presented here at the  $1 \sigma$
level.  It  is extremely reassuring that two  wildly different methods
(CLF modeling versus bispectrum analysis) yield values for the bias in
such good agreement.

Finally,  upon combining  the priors  on $\beta$  and  $\langle M_{\rm
  vir}/L \rangle_{\rm  cl}$ we  obtain the 68  and 95  percent contour
levels  indicated by dotted  contours.  Marginalizing  over $\sigma_8$
and  $\Omega_m$  we   obtain  $\Omega_m  =  0.27^{+0.14}_{-0.10}$  and
$\sigma_8  =  0.77^{+0.10}_{-0.14}$  (both  95\%  CL),  respectively.  
Whereas the constraint on  $\Omega_m$ is perfectly consistent with the
concordance value, the constraint  on the cluster mass-to-light ratios
strongly    favors   low-$\sigma_8$    models   over    the   standard
$\sigma_8=0.9$.  Keep   in  mind,  however,  that   this  analysis  is
restricted to flat  $\Lambda$CDM cosmologies with $\Omega_b h^2=0.02$,
$h=0.7$, and $n_s=1.0$. In what follows we relax these constraints and
perform  a more  detailed analysis  including additional  data  on CMB
anisotropies.

\section{A joint CMB plus LSS analysis}
\label{sec:joint}

\subsection{Cosmic Microwave Background}
\label{sec:cmb}

Undoubtedly the most  stringent constraints on cosmological parameters
come from measurements of  the angular CMB temperature power spectrum.
Results from  COBE (Bennett \etal 1996),  BOOMERANG (Netterfield \etal
2002), MAXIMA  (Hanany \etal 2000),  DASI (Halverson \etal  2002), VSA
(Scott \etal 2002),  CBI (Pearson \etal 2002) and  WMAP (Bennett \etal
2003) have  combined to produce  a power  spectrum of  CMB temperature
anisotropies with  the first three acoustic peaks  resolved.  This has
resulted  in stringent constraints  on cosmological  parameters (e.g.,
Wang,  Tegmark \&  Zaldarriaga 2002;  de Bernardis  \etal  2002; Pryke
\etal 2002;  Sievers \etal 2002; Efstathiou \etal  2002; Spergel \etal
2003).   Unfortunately,  because of  the  unknown  amount of  Thompson
scattering experienced by CMB photons, these data only put constraints
on  the combination $\sigma_8  {\rm e}^{-\tau}$.   Here $\tau$  is the
optical depth for Thompson  scattering, which depends, amongst others,
on the  reionization redshift  $z_{\rm re}$.  The  absence of  a clear
Gunn-Peterson trough  in the spectra  of high-redshift quasars  puts a
lower limit on the reionization  epoch of $z_{\rm re} \gta 5.5$ (e.g.,
Becker  \etal 2001;  Fan \etal  2001),  and therewith  on $\tau$.   In
addition,  the observed  height of  the first  acoustic  peak requires
$\tau \lta 0.4$  (e.g., Griffiths, Barbosa \& Liddle  1999; Ruhl \etal
2002).   Because  of  these   limits  on  $\tau$  the  uncertainty  on
$\sigma_8$ is limited to about 30 percent.

To  illustrate   the  constraints  that  pre-WMAP  CMB   data  put  on
($\Omega_m$,$\sigma_8$)-parameter space we  use the Monte-Carlo Markov
(MCM) chains of Lewis \& Bridle (2002; hereafter LB02).  These consist
of   large  numbers  of   cosmological  models,   each  of   which  is
characterized  by $N$  model parameters.   The chains  are constructed
using the Metropolis-Hastings algorithm, which ensures that the number
density distribution of these  models in the $N$-dimensional parameter
space  traces  out the  posterior  probability  distribution that  the
observed  data  originates from  the  given  cosmology  (see LB02  and
references therein).  The clear  advantage of this MCM chain technique
is  that it samples  the entire  $N$-dimensional posterior  giving far
more information  than just the marginalized distributions  of each of
the $N$  individual parameters (which is what  is commonly presented).
In  addition,  it  is  straightforward  to  extract  the  marginalized
distribution  in   any  sub-dimensional  parameter   space  by  simple
projection.

Here  we  use  the  publicly  available $N=6$  Markov  Chain  of  LB02
consisting of $\sim 11500$  weakly correlated cosmological models. The
six parameters  that label each  model are: $\Omega_b  h^2$, $\Omega_c
h^2$, $h$, $n_s$, the redshift  $z_{\rm re}$ at which the reionization
fraction is  a half,  and $A_s$,  a measure for  the amplitude  of the
initial  power spectrum.   Most of  these parameters  have  weak, flat
priors (see LB02  for details), except for the  Hubble parameter which
has a  Gaussian prior  of $h=0.72  \pm 0.08$ (taken  from the  HST key
project,  Freedman  \etal 2001).   All  models  are computed  assuming
purely  adiabatic,  scalar-only,  Gaussian  primordial  perturbations,
$\Omega_K  = 0$,  non-interacting CDM,  and an  effective  equation of
state  for  the  dark  energy  $w  =  -1$  (corresponding  to  a  pure
cosmological  constant).   The  data  used  to  construct  the  chains
consists of a  combination of the results of  COBE, BOOMERANG, MAXIMA,
DASI,  VSA, and  CBI  in the  form  of band  power  estimates for  the
temperature CMB power spectrum. The CBI data with $l > 2000$, which is
most  likely due  to non-linear  effects,  is ignored.   

\begin{figure*}
\centerline{\psfig{figure=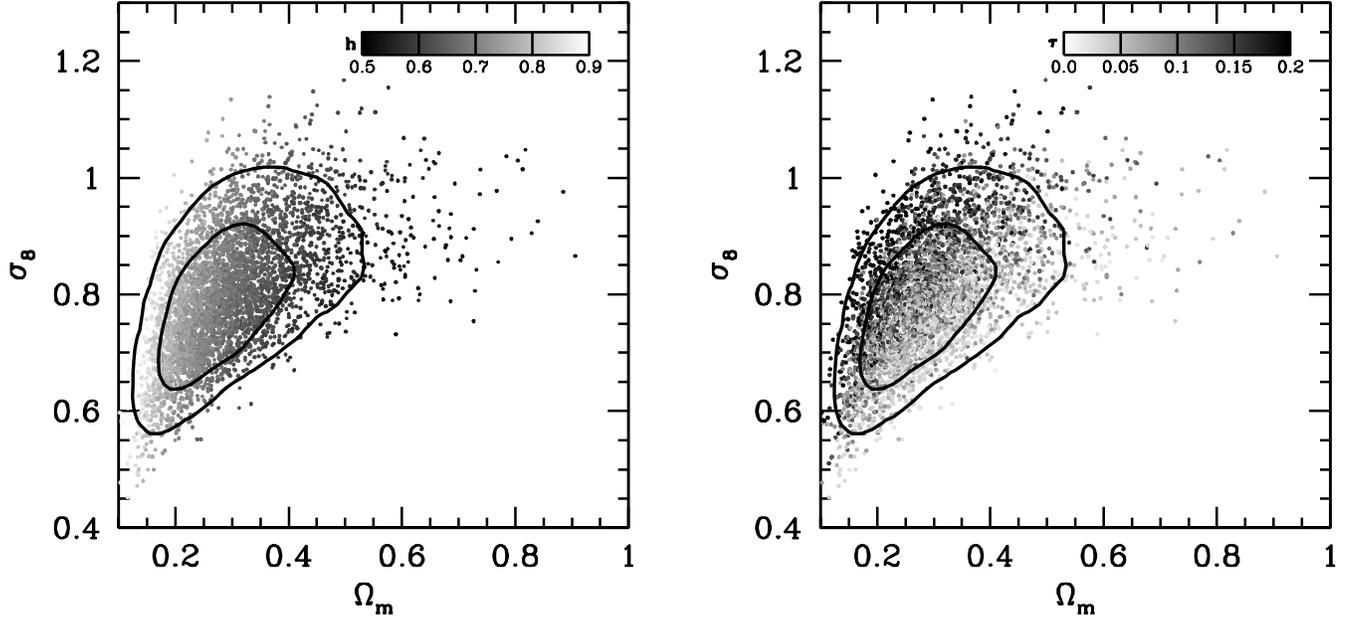,width=\hdsize}}
\caption{The  points  correspond   to  cosmological  models  from  the
Monte-Carlo Markov  chain of LB02, and are  gray-scale coded according
to their  values for $h$ (left  panel) and $\tau$  (right panel).  The
chain  is  constructed such  that  the  number  density of  points  is
proportional   to   the   marginalized   likelihood   ${\cal   L}_{\rm
CMB}(\Omega_m,\sigma_8)$ with the 68  and 95 percent confidence levels
indicates with contours. See text for details.}
\label{fig:cmb}
\end{figure*}

Figure~\ref{fig:cmb} plots the distribution of models of the MCM chain
in  the  ($\Omega_m$, $\sigma_8$)  parameter  space, gray-scale  coded
according  to  their  value  of  $h$ (left  panel)  or  $\tau$  (right
panel). As mentioned above,  the number density distribution of models
is directly proportional to the marginalized likelihood ${\cal L}_{\rm
CMB}(\Omega_m,\sigma_8)$.   The two  contours indicate  the 68  and 95
percent   confidence  levels,  obtained   by  integrating   under  the
likelihood surface.   Marginalizing over $\sigma_8$  and $\Omega_m$ we
obtain   $\Omega_m    =   0.26^{+0.22}_{-0.12}$   and    $\sigma_8   =
0.80^{+0.19}_{-0.18}$  (95\%   CL),  respectively.   The   95  percent
confidence interval  for $\sigma_8$ therefore covers  the entire range
of values for  $\sigma_8$ quoted in the literature.   As apparent from
the gray-scale coding  in the panel on the  right, this uncertainty on
$\sigma_8$  is  due to  the  $\sigma_8$--${\rm e}^{-\tau}$  degeneracy
mentioned above.   Higher values of $\sigma_8$  (for given $\Omega_m$)
correspond to higher optical depth, and therefore to a higher redshift
of reionization.   The constraints  on $\Omega_m$ are  almost entirely
due to the  Gaussian prior on $h$, as is  immediately evident from the
gray-scale  coding in  the left  panel.   Whereas the  CMB data  alone
mainly  constrains  $\Omega_K$  (through  the location  of  the  first
acoustic peak), the combination of CMB data plus a prior on the Hubble
constant  already   puts  useful  constraints   on  $\Omega_m$  (e.g.,
Rubi\~no-Martin \etal 2002; LB02).

\subsection{Combining CMB plus 2dFGRS}
\label{sec:combine}
 
An  extremely attractive  feature  of the  MCM  chains is  that it  is
straightforward to compute likelihoods including additional data.  The
weight of each model is simply adjusted proportional to the likelihood
under  the new constraint,  a technique  known as  importance sampling
(see  LB02   for  details).   The  new,   marginalized  likelihood  is
subsequently obtained from the {\it weighted} number density of points
as a function of $\Omega_m$ and $\sigma_8$.  Using importance sampling
we compute the combined likelihood  ${\cal L}_{\rm tot}$ that both the
CMB data plus the $\Phi(L)$ and $\xi_{\rm gg}(r,L)$ originate from the
given cosmology.

\begin{figure*}
\centerline{\psfig{figure=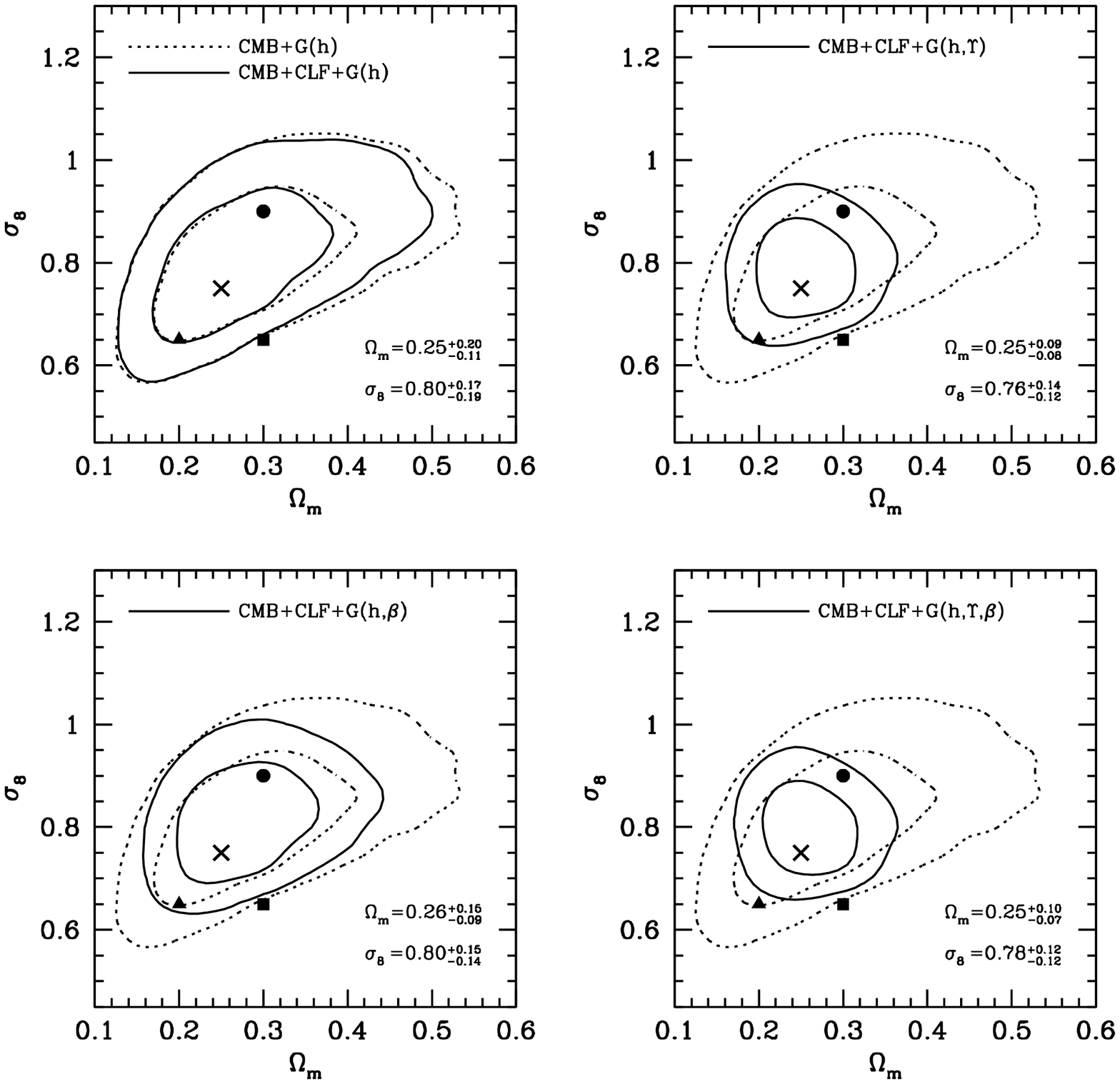,width=\hdsize}}
\caption{The dotted contours  in each of the four  panels indicate the
68 and  95 percent confidence  levels on $\Omega_m$ and  $\sigma_8$ as
determined from  the CMB  data plus the  Gaussian prior $G(h)$  on the
Hubble parameter:  $h=0.72 \pm 0.08$. The solid  contours indicate the
same confidence levels of the combined likelihood ${\cal L}_{\rm tot}$
with  the priors  as  indicated,  using the  same  nomenclature as  in
Figure~\ref{fig:beta}.}
\label{fig:combi}
\end{figure*}

The   68  and  95   percent  confidence   levels  of   ${\cal  L}_{\rm
tot}(\Omega_m,\sigma_8)$  are  shown  in   the  upper  left  panel  of
Figure~\ref{fig:combi}   (solid  lines).   As   is  apparent   from  a
comparison  with  the  likelihood  from  the CMB  data  alone  (dotted
contours), the LSS  data adds virtually no new  constraints.  In other
words, any  model (in the 6-dimensional parameter  space studied here)
that fits  the CMB  data also fits  the galaxy correlation  lengths as
function  of luminosity.   Although  the lack  of  improvement on  the
constraints  is  perhaps somewhat  disappointing,  it  is a  beautiful
demonstration  of  the  level   of  concordance  among  two  completely
independent sets of data.

As in  Section~\ref{sec:CLF} above we  can tighten the  constraints on
$\Omega_m$ and $\sigma_8$ by using priors on the cluster mass-to-light
ratio and/or  $\beta$.  Including  the Gaussian prior  $\langle M_{\rm
  vir}/L \rangle_{\rm cl}  = (350 \pm 70) h \MLsun$  results in the 68
and  95 percent  confidence levels  shown in  the upper  right  panel. 
Clearly, the  constraints on $\langle M_{\rm  vir}/L \rangle_{\rm cl}$
restrict  both  $\Omega_m$ and  $\sigma_m$  to  relatively low  values
(compared  to  the   parameter  space  allowed  by  the   CMB  data).  
Marginalizing  over $\sigma_8$  and $\Omega_m$  we obtain  $\Omega_m =
0.25^{+0.09}_{-0.08}$ and $\sigma_8 = 0.76^{+0.14}_{-0.12}$ (both 95\%
CL), respectively.  This is inconsistent with the standard $\sigma_8 =
0.9$  at $\sim  2 \sigma$.   Fairly  similar results  are obtained  if
instead of  the prior on  $\langle M_{\rm vir}/L \rangle_{\rm  cl}$ we
include  the Gaussian  prior  $\beta  = 0.49  \pm  0.09$ ($\Omega_m  =
0.26^{+0.15}_{-0.09}$,  $\sigma_8 =  0.80^{+0.15}_{-0.14}$,  see lower
left panel  of Figure~\ref{fig:combi}).  Finally,  upon including both
priors we obtain the confidence  levels shown in the lower right panel
with    marginalized    probability    distributions    $\Omega_m    =
0.25^{+0.10}_{-0.07}$ and $\sigma_8 = 0.78 \pm 0.12$ (both 95\% CL).

In  summary,  the observed  clustering properties of  galaxies  are in
perfect agreement  with matter power spectra  that fit the current CMB
data. The conditional luminosity function  models presented here allow
us to compute mass-to-light  ratios and galaxy   bias in a  completely
self-consistent way, which in turn  allows us to significantly tighten
the  constraints   on   $\Omega_m$ and    $\sigma_8$.   The  strongest
constraints  come from the  observed mass-to-light  ratio of clusters,
which strongly   argues for   cosmologies  with  both $\Omega_m$   and
$\sigma_8$ reduced  by $\sim 15$ percent with  respect to the standard
values of $0.3$ and $0.9$, respectively.
\begin{figure*}
\centerline{\psfig{figure=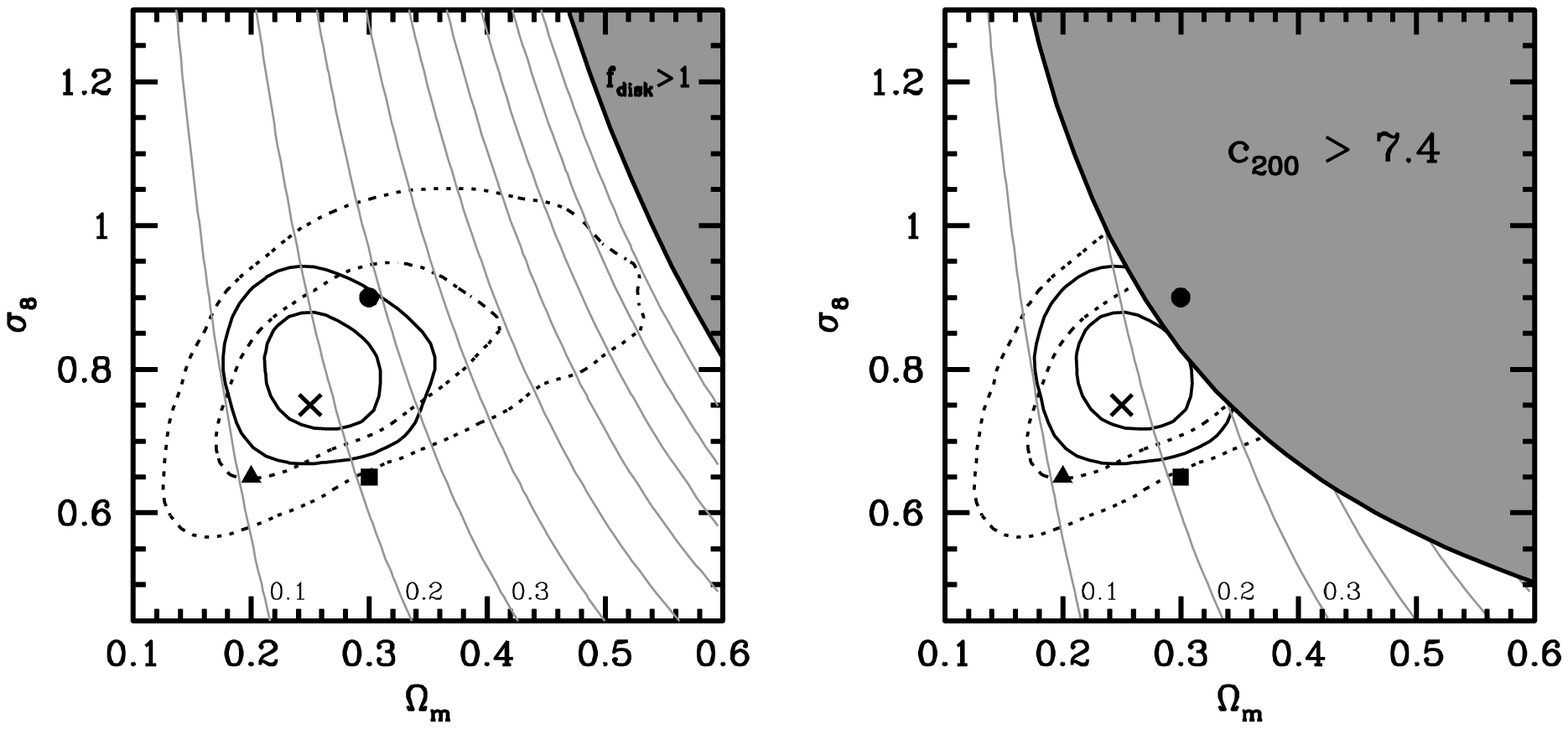,width=\hdsize}}
\caption{Thin gray lines in  both panels indicate contours of constant
$f_{\rm disk}$ (equation~[\ref{fdisk}]). The contours corresponding to
$f_{\rm disk} = 0.1$, $0.2$,  and $0.3$ are labelled for clarity.  The
gray areas correspond to $f_{\rm  disk} > 1$ (left panel) and $c_{200}
> 7.4$ (right panel), and indicate the regions of parameter space that
are excluded according to  the observed baryonic Tully-Fisher relation
and the  observed rotation curves  of LSB galaxies,  respectively (see
text for details).   The dotted contours in both  panels correspond to
the   68  and  95   percent  confidence   levels  of   ${\cal  L}_{\rm
CMB}(\Omega_m,\sigma_8)$          (cf.           Figures~\ref{fig:cmb}
and~\ref{fig:combi}),  while  the  solid  contours indicate  the  same
levels of  confidence but for  ${\cal L}_{\rm tot}(\Omega_m,\sigma_8)$
with the Gaussian priors $\langle M_{\rm vir}/L \rangle_{\rm cl}= (350
\pm 70) h \MLsun$ and $\beta  = 0.49 \pm 0.09$ (cf.  lower right panel
of    Figure~\ref{fig:combi}).      The    symbols    are     as    in
Figure~\ref{fig:clf} and correspond to the four models in Table~1.}
\label{fig:tf}
\end{figure*}

\section{Implications}
\label{sec:implic}

As outlined in  Section~\ref{sec:intro}, despite the large uncertainty
in $\sigma_8$, most numerical  simulations of structure formation in a
$\Lambda$CDM cosmology have adopted $\Omega_m=0.3$ and $\sigma_8=0.9$.
This  has also  been the  preferred  cosmology for  studies of  galaxy
formation. Yet,  several studies over  the past years,  including this
one, have  suggested values for $\Omega_m$ and/or  $\sigma_8$ that are
reduced with respect to these standard values.

In   this  section  we  investigate    the   implications that   small
modifications of $\sigma_8$ and $\Omega_m$  have on the structure  and
formation of galaxies and  their associated CDM haloes. In particular,
we focus on two problems that have been identified in recent years for
the standard $\Lambda$CDM  cosmology;  the problem of matching  the TF
zero-point and the   inconsistencies  between observed and   predicted
rotation curves for dwarf and low surface brightness galaxies.

\subsection{The Baryonic Tully-Fisher Relation}
\label{sec:TFbar}

Disk  galaxies  follow  a scaling  relation   between  luminosity  and
rotation velocity known as the Tully-Fisher  (TF) relation.  Since the
rotation velocity is  a dynamical mass measure,  the zero-point of the
Tully-Fisher relation  sets  a characteristic  mass-to-light ratio and
can therefore,   in  principle,  be  used to   constrain  cosmological
parameters  (see e.g.,  van den Bosch  2000).    Here we focus  on the
so-called ``baryonic'' Tully-Fisher relation  between disk  {\it mass}
and  rotation  velocity (McGaugh  \etal 2000;  Bell  \& de Jong 2001).
Observationally,   the disk mass is  obtained  by multiplying the disk
luminosity  with  the stellar    mass-to-light  ratio and   adding the
contribution of the cold gas  in the disk.  Here we use the results of
McGaugh \etal (2000)\footnote{We have checked that our results are not
significantly different if we  use the somewhat different  baryonic TF
relation of Bell \& de Jong (2001).}, who found
\begin{equation}
\label{barTF}
M_{\rm disk} = 1.97 \times 10^{9} h^{-2} \Msun \,
\left( {V_{\rm rot} \over 100 \kms} \right)^4
\end{equation}

Defining dark matter haloes as  spheres with an average density inside
the virial radius, $r_{\rm vir}$, that is $\Delta_{\rm vir}$ times the
critical density $\rho_{\rm crit}$, one obtains the following relation
between  the  virial mass  $M_{\rm  vir}$  and  the circular  velocity
$V_{\rm vir}$ at the virial radius:
\begin{equation}
\label{Vvir}
M_{\rm vir} = 2.33 \times 10^{11} h^{-1} \Msun 
\left( {V_{\rm vir} \over 100 \kms} \right)^3 \,
\left( {\Delta_{\rm vir} \over 200} \right)^{-1/2}
\end{equation}
The   cosmology    dependence  enters    through   $\Delta_{\rm  vir}$
(eq.~[\ref{dvir}]).  The mass of a disk  galaxy that forms inside this
halo can be expressed as
\begin{equation}
\label{Mdisk}
M_{\rm disk} = f_{\rm disk} \, \left( {\Omega_b \over \Omega_m} \right)
\, M_{\rm vir}
\end{equation}
Here $f_{\rm disk}$  is the fraction  of baryonic material in the halo
that   ends up  in the  disk.  Equating~(\ref{Mdisk})  to the observed
baryonic TF relation~(\ref{barTF}) yields
\begin{eqnarray}
\label{fd}
f_{\rm disk} & = & 0.423 \, h \, \Omega_m \, 
\left( {\Omega_b h^2 \over 0.02} \right)^{-1} \, 
\left( {V_{\rm rot} \over V_{\rm vir}} \right)^4 \times \nonumber \\
& & \left( {\Delta_{\rm vir} \over 200} \right)^{1/2}
\left( {V_{\rm vir} \over 100 \kms} \right)
\end{eqnarray}
This fraction has to  obey $f_{\rm disk} \leq 1$,  at least  under the
standard assumption that the   baryonic  fraction inside  dark  matter
haloes can not exceed the universal value. The strongest constraint on
cosmological  parameters     comes   from   the   high-$V_{\rm   vir}$
end. Using~(\ref{Vvir}) and  evaluating $f_{\rm disk}$ at $M_{\rm vir}
=  3 \times 10^{12}  h^{-1} \Msun$, which is  roughly the maximum halo
mass for disk galaxies, one obtains
\begin{equation}
\label{fdisk}
f_{\rm disk} = 0.991 \, h \Omega_m \, 
\left( {\Omega_b h^2 \over 0.02} \right)^{-1} \, 
\left( {V_{\rm rot} \over V_{\rm vir}} \right)^4 \,
\left( {\Delta_{\rm vir} \over 200} \right)^{2/3}
\end{equation}

This fraction depends  strongly on the ratio of  the observed rotation
velocity to  the circular  velocity of the  halo. In the  CDM paradigm
dark  matter  haloes follow  the  universal  NFW density  distribution
\begin{equation}
\label{NFWprof}
\rho(r) = {\bar{\delta} \, \bar{\rho} \over (r/r_s) (1 + r/r_s)}
\end{equation}
(Navarro, Frenk \& White 1997). Here $r_s$ is a characteristic radius,
$\bar{\rho} = \Omega_m \rho_{\rm crit}$  is the average density of the
Universe, and $\bar{\delta}$ is a dimensionless amplitude which can be
expressed  in terms  of  the halo  concentration  parameter $c  \equiv
r_{\rm vir}/r_s$ as
\begin{equation}
\label{avdelta}
\bar{\delta} = {\Delta_{\rm vir} \over 3} \, {c^3 \over {\rm ln}(1+c)
- c/(1+c)}
\end{equation}
The circular  velocity curve of  a NFW density distribution  reaches a
maximum  $V_{\rm max}$  at a  radius $r  = 2.163  r_s$.  The  ratio of
$V_{\rm max}$ to the virial velocity is given by
\begin{equation}
\label{vrat}
{V_{\rm max} \over V_{\rm vir}} = 0.465 \,
\sqrt{c \over {\rm ln}(1+c) - c/(1+c)}
\end{equation}
and is typically  larger than unity. We use the  model of Eke, Navarro
\&  Steinmetz (2001)  to  compute $c$  as  function of  halo mass  and
cosmology, and we assume that  $V_{\rm rot} = V_{\rm max}$, i.e., that
the observed, flat  part of the rotation curve  coincides with $V_{\rm
max}$.  Detailed  models (Mo, Mao \&  White 1998; van  den Bosch 2001;
2002) indicate that in cases were $f_{\rm disk}$ is sufficiently large
$V_{\rm rot} > V_{\rm max}$ due to the contribution of the disk to the
circular velocity. We  ignore this effect for the  moment, but caution
that the $f_{\rm disk}$ thus derived may be an underestimate.

In Figure~\ref{fig:tf} we plot  contours of constant $f_{\rm disk}$ as
function of  $\Omega_m$ and $\sigma_8$.   The gray region in  the left
panel indicates the cosmologies for which $f_{\rm disk} > 1$ and which
are  thus  excluded.  In  addition,  we plot  the  68  and 95  percent
confidence  levels of ${\cal  L}_{\rm CMB}$  (dotted contours)  and of
${\cal   L}_{\rm  tot}$   plus  priors   on  $\langle   M_{\rm  vir}/L
\rangle_{\rm cl}$  and $\beta$ (solid contours).   The constraint from
the TF zero-point ($f_{\rm disk}  \leq 1$) only rules out models which
are already  inconsistent with  the CMB data  at more than  95 percent
confidence.   Typically  $f_{\rm   disk}$  decreases  with  decreasing
$\Omega_m$  and decreasing  $\sigma_8$.   Note that  for the  standard
$\Lambda$CDM cosmology,  with $\Omega_m=0.3$ and  $\sigma_8=0.9$, only
about  30 percent  of the  available baryons  in a  $3  \times 10^{12}
h^{-1} \Msun$  halo should  be present  in the disk.   In the  case of
cosmologies  favored by  the CMB  plus LSS  data presented  here, this
fraction is even lower, between 10  and 20 percent.  Thus, in order to
explain  the zero-point  of the  baryonic  TF relation,  only a  minor
fraction of the baryonic mass  within $r_{\rm vir}$ can become part of
the disk; the  remaining baryonic material either remains  in the halo
as hot  gas, or is  expelled from the  halo. Note that  $f_{\rm disk}$
scales with  $V_{\rm vir}$  such that this  fraction is even  lower in
lower mass haloes.  This requirement for a physical mechanism that can
prevent baryons  from ending up in  cold gas or stars  is also obvious
from a  direct comparison  of the halo  mass function with  the galaxy
luminosity function  (e.g., White \&  Rees 1978; White \&  Frenk 1991;
YMB03)  and  remains  one  of  the most  challenging  puzzles  in  the
framework of galaxy formation.
\begin{figure*}
\centerline{\psfig{figure=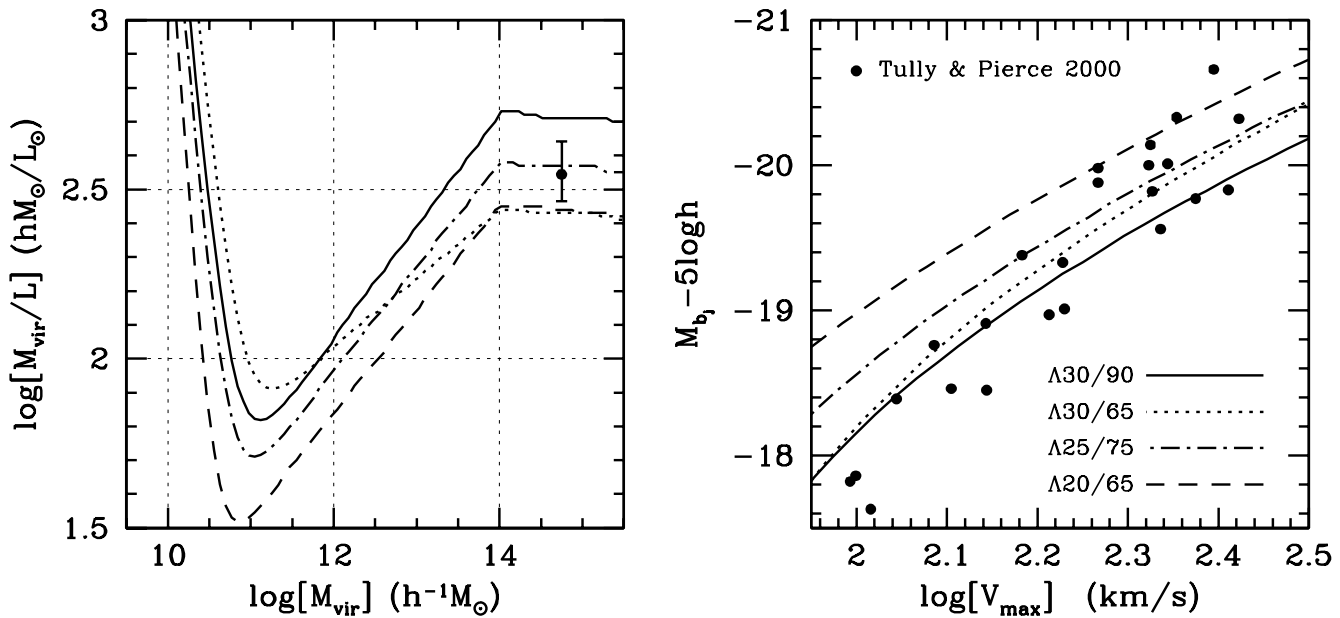,width=\hdsize}}
\caption{{\it Left panel:} The average mass-to-light ratio as function
of halo  mass for the four  cosmological models listed  in Table~1, as
obtained from the  best-fit CLF.  The solid dot  indicates the average
mass-to-light  ratio of clusters  $\langle M_{\rm  vir}/L \rangle_{\rm
cl} = 350 \pm 70 h \MLsun$, where the errorbars indicate the $1\sigma$
interval.   {\it Right  panel:}  Tully-Fisher relations  for the  four
cosmologies  listed  in Table~1,  obtained  from  the $\langle  M_{\rm
vir}/L \rangle(M)$ shown  in the left panel as  described in the text.
For comparison, the solid  dots indicate the Tully-Fisher relation for
the `local calibrator sample' of Tully \& Pierce (2000).}
\label{fig:ml}
\end{figure*}

\subsection{The Tully-Fisher Zero-Point}
\label{sec:tulfis}

Detailed semi-analytical models for the  formation of galaxies have so
far been unsuccessful in simultaneously fitting the  galaxy LF and the
TF zero-point.  Initially, when these  studies focussed on Einstein-de
Sitter cosmologies,  the  discrepancies  where found to   be extremely
large,  with models  tuned  to fit  the LF  predicting much too  faint
luminosities for a given rotation  velocity (e.g., Kauffmann, White \&
Guiderdoni 1993; Cole \etal 1994).  Heyl \etal (1995) showed that this
could be  significantly improved upon  by lowering $\Omega_m$, but the
overall agreement   remained unsatisfactory.  Even  for  the currently
popular  $\Lambda$CDM cosmology with $\Omega_m=0.3$ and $\sigma_8=0.9$
no  semi-analytical  models  presented   to  date   has been able   to
simultaneously  fit the  LF and TF   zero-point (Somerville \& Primack
1999; Cole  \etal 2000; Benson  \etal 2000,  2002; Mathis \etal 2002).
Remarkably enough, YMB03 found exactly the same problem when comparing
the average   mass-to-light ratios inferred from    the CLF with those
required  to fit  the  TF zero-point. Since the   CLF method makes  no
assumptions about how  galaxies form, this  indicates that the problem
is not related to the (poorly understood) physics of galaxy formation,
but rather is a problem of more fundamental, cosmological origin.

As  shown  in   Section~\ref{sec:TFbar},  the  TF  zero-point  depends
strongly on the concentration of dark matter haloes: more concentrated
haloes  have a  higher $V_{\rm  max}/V_{\rm vir}$,  and thus  a higher
rotation   velocity  for   a  given   disk  luminosity.    Since  halo
concentrations  are strongly  cosmology dependent,  the  TF zero-point
problem  outlined   above  may  simply  indicate   that  the  standard
cosmological  concordance parameters  are  not correct.   In order  to
access  the impact  of  small changes  in  $\Omega_m$ and  $\sigma_8$,
consider the  four cosmologies listed  in Table~1.  The left  panel of
Figure~\ref{fig:ml} plots  the average mass-to-light  ratios, $\langle
M_{\rm  vir}/L \rangle(M_{\rm  vir})$, as  obtained from  the best-fit
CLFs.  The general  trend is the same for all  models: at $M_{\rm vir}
\lta 10^{11} h^{-1} \Msun$  the mass-to-light ratios strongly increase
with decreasing halo mass.  This is required in order to reconcile the
faint slope of the galaxy  LF with the relatively steep low-mass slope
of the halo mass function  (cf., YMB03).  At $M_{\rm vir} \geq 10^{14}
h^{-1}  \Msun$ the mass-to-light  ratio $\langle  M/L \rangle$  is, by
construction  (see Appendix~A) constant  at $\langle  M/L \rangle_{\rm
cl}$, in good agreement  with the observed flattening of mass-to-light
ratios with  increasing scale (Bahcall \etal  1995, 2000).  Typically,
lowering $\sigma_8$  (while keeping all  other cosmological parameters
fixed)  reduces  $\langle  M/L  \rangle_{\rm cl}$  and  increases  the
mass-to-light ratios on the scales  of galaxies (cf.  models \lamA and
\lamB).  Simultaneously  reducing $\sigma_8$ and  $\Omega_m$ along the
valley floor of ${\cal L}_{\rm CLF}$, however, leads to a reduction of
$\langle M_{\rm vir}/L \rangle$ on  all mass scales (cf. models \lamA,
\lamC, and \lamD).

We can  use the  CLFs to  compute predictions for  the TF  relation as
follows. We assume that TF disk galaxies are the brightest galaxies in
their  haloes.  From  the CLF,  it is  straightforward to  compute the
average luminosity of the  brightest galaxy, $\langle L_c \rangle$, as
function of halo mass (see Appendix~A), which we convert to magnitudes
in  the photometric $b_J$-band  using $M_{\odot,  b_J}=5.3$.  Finally,
the  maximum   rotation  velocity  $V_{\rm  max}$   is  obtained  from
equation~(\ref{vrat}).  Assuming that these $V_{\rm max}$ are equal to
the TF  rotation velocities, we obtain  the TF relations  shown in the
right  panel of  Figure~\ref{fig:ml}.   For comparison,  we also  plot
(solid circles)  the $b_J$ band  TF relation of the  `local calibrator
sample' of  Tully \&  Pierce (2000). Here  we have  converted $B$-band
magnitudes to the $b_J$ band using $b_j = B - 0.28 \, (B-V)$ (Blair \&
Gilmore 1982) and adopting $B-V=0.7$, which corresponds roughly to the
average color of disk galaxies (de Jong 1996).

Compared to  the data, the  TF relation for the  standard $\Lambda$CDM
cosmology \lamA  is somewhat too shallow,  clearly underpredicting the
luminosity of the more massive disk galaxies. Since the CLFs are tuned
to fit  the observed  LF, this illustrates  the TF  zero-point problem
outlined above.  Model \lamB,  however, predicts a somewhat steeper TF
relation, in  better agreement with  the data, while models  \lamC and
\lamD  predict TF  zero-points that  are brighter  than that  of model
\lamA by as  much as almost an entire  magnitude.  Clearly, relatively
small  changes in  $\Omega_m$ and/or  $\sigma_8$ with  respect  to the
standard concordance values of $0.3$ and $0.9$, respectively, strongly
alleviate the problem (see  also Seljak 2002b,c).  Although a definite
answer  requires  a  more  thorough  investigation,  it  is  extremely
encouraging that the same cosmological  model that is preferred by the
observed abundance and  clustering of galaxies, predicts mass-to-light
ratios  on   galactic  scales  that  better  match   the  observed  TF
zero-point.   Remarkably enough,  as shown  in Section~\ref{sec:TFbar}
above, this  cosmology also requires  lower values of  $f_{\rm disk}$,
and thus a stronger efficiency of preventing baryons from ending up in
the disk.   It remains to be  seen whether detailed  models for galaxy
formation, can simultaneously match the  LF and the TF zero-point when
considering cosmologies such as \lamC.

\subsection{Rotation Curves}
\label{sec:RC}

In the past years numerous  authors have pointed out that the rotation
curves  of dwarf  and low-surface  brightness (LSB)  galaxies indicate
dark matter haloes that are less centrally concentrated than predicted
by  the  standard $\Lambda$CDM  cosmology  with $\sigma_8=0.9$  (e.g.,
Moore  1994; Burkert  1995; van  den  Bosch \etal  2000; Borriello  \&
Salucci 2001; Blais-Ouelette, Amram \& Carignan 2001; de Blok, McGaugh
\& Rubin 2001; de Blok \& Bosma 2002; Swaters \etal 2003).
\begin{figure}
\centerline{\psfig{figure=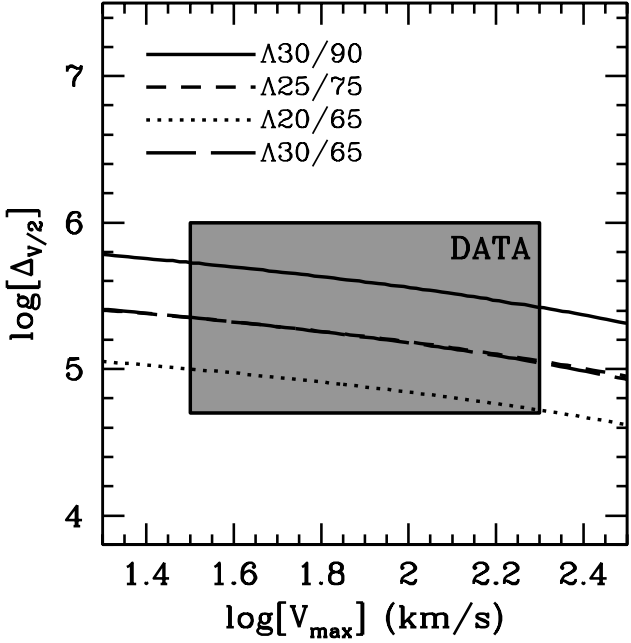,width=\hssize}}
\caption{The  concentration of dark  matter haloes,  expressed through
the  dimensionless  quantity  $\Delta_{V/2}$ (equation~[\ref{dv}])  as
function of the maximum circular velocity of the halo. Predictions are
shown for the four cosmological models listed in Table~1 as indicated,
and  are  computed   using  the  Eke  \etal  (2001)   model  for  halo
concentrations.  The  gray area  labelled `DATA' indicates  the region
occupied  by  observed  galaxies  (see  Alam  \etal  2002).   Reducing
$\Omega_m$  and/or  $\sigma_8$ reduces  halo  concentrations and  thus
$\Delta_{V/2}(V_{\rm max})$. Note that the $\Delta_{V/2}(V_{\rm max})$
of cosmologies \lamB and~\lamC almost exactly overlap.}
\label{fig:rc}
\end{figure}

Since the average  concentration of a halo of  given mass is cosmology
dependent, a constraint on  halo concentrations translates into one on
cosmological parameters.  This principle was recently used by McGaugh,
Barker  \&   de  Blok  (2003).   Using  the   high  resolution  hybrid
H$\alpha$-HI  rotation curves of  LSB galaxies  presented by  de Blok,
McGaugh \&  Rubin (2001) and de  Blok \& Bosma (2002)  they derived an
upper limit  for the  mean observed halo  concentration of  $c_{200} <
7.4$.  Here $c_{200}=r_{200}/r_s$ with  $r_{200}$ the radius inside of
which  the average density  of the  halo is  $200$ times  the critical
density.  Using the model outlined  in Navarro \etal (1997) to compute
$c_{200}$  for given  cosmology  and halo  mass  McGaugh \etal  (2003)
obtain that
\begin{equation}
\label{cmax}
\sigma_8 \Gamma_{0.6} < 0.23
\end{equation}
with
\begin{equation}
\label{gam6}
\Gamma_{0.6} = \Omega_m^{0.6} \, h \, {\rm exp}(-\Omega_b - \sqrt{2h}
\Omega_b/\Omega_m) - 0.32 \, (n_s^{-1} - 1)
\end{equation}
This is based  on the assumption that all LSB  galaxies reside in dark
matter haloes with $M_{200} =  10^{12} h^{-1} \Msun$. Since in general
one  expects  that $M_{200}$  is  smaller  than  this, and  since  $c$
decreases  with increasing  halo mass,  this is  a  conservative upper
limit.

The gray  region  in the   upper  right panel of   Figure~\ref{fig:tf}
indicates  the  part of ($\Omega_m$,  $\sigma_8$)  parameter space for
which haloes with $M_{200} = 10^{12} h^{-1} \Msun$ have concentrations
$c_{200} > 7.4$. Here we have  adopted $\Omega_b h^2 = 0.02$, $h=0.7$,
and  $n_s=1.0$.    As is  apparent, the  constraint~(\ref{cmax})  puts
stringent constraints on $\Omega_m$  and $\sigma_8$. In fact, if taken
at face value, a large fraction of the cosmologies that are consistent
with the CMB   data at the 95   percent confidence level or  better is
ruled out,    including the standard   $\Lambda$CDM  cosmology  \lamA.
However, virtually  the  entire 68 percent confidence  region obtained
from   the    joined    CMB   plus   LSS   analysis     presented   in
Section~\ref{sec:joint}   above is {\it consistent} with~(\ref{cmax}).
Thus, as    with the TF  zero-point,  the  problem with   the observed
rotation curves of dwarf and LSB galaxies may  simply be alleviated by
adopting a cosmology with somewhat  lower $\sigma_8$ and/or $\Omega_m$
(see also Zentner \& Bullock 2002).

Unfortunately,  the  robustness of   this  result is  questionable for
several reasons. First of all, we have adopted a halo mass of $M_{200}
=  10^{12} h^{-1} \Msun$, whereas  in reality LSB galaxies will reside
in haloes with a variety of halo masses. Secondly,  in order to obtain
a measure of $c$ from an observed  rotation curve one generally fits a
mass model  to the data.   In some cases, however, no  good fit can be
obtained for any value of $c$. This is often interpreted as indicating
that dark matter haloes  do not follow an NFW  profile (i.e., the dark
matter    is  not  cold and  collisionless),    but may  also indicate
non-circular   motions or other distortions   (see e.g., Salucci 2001;
Swaters \etal 2003).

A more  robust comparison of models  with data was  suggested by Alam,
Bullock \& Weinberg (2002).   Rather than using the halo concentration
parameter $c$, which is difficult to extract from an observed rotation
curve,  Alam  \etal   (2002)  introduced  the  dimensionless  quantity
\begin{equation}
\label{dv}
\Delta_{V/2} = {\bar{\rho}(R_{V/2}) \over \rho_{\rm crit}} = 
{1 \over 2} \left( {V_{\rm max} \over H_0 R_{V/2}} \right)^2
\end{equation}
as a    more robust  measure  of   the concentration of    dark matter
haloes. Here $R_{V/2}$ is defined as the  radius at which the rotation
curve falls  to half  of   its  maximum   value  $V_{\rm max}$,    and
$\Delta_{V/2}$  thus  measures the  mean  dark  matter  density inside
$R_{V/2}$ in units of  the critical density $\rho_{\rm  crit}$. Unlike
$c$ and $V_{\rm vir}$, the parameters $\Delta_{V/2}$ and $V_{\rm max}$
are easy to obtain from an  observed rotation curve, without having to
resort  to mass model fitting. Under  the assumption that the observed
rotation curve is dominated by the  contribution of the dark matter, a
valid  assumption in   the case of   LSB galaxies,  $V_{\rm  max}$ and
$R_{V/2}$ can be converted directly to $c$ and  $V_{\rm vir}$ and thus
be compared to cosmology dependent predictions.

For the  four cosmologies listed in Table~1  we compute $\Delta_{V/2}$
as function  of $V_{\rm max}$ as  follows.  We assume  that CDM haloes
have a NFW density distribution and that halo concentrations depend on
halo mass  and cosmology according to  the model of  Eke \etal (2001).
$V_{\rm   max}$  then   follows   from  equation~(\ref{vrat}),   while
$\Delta_{V/2}$   is  related  to   the  concentration   parameter  $c$
as\footnote{This relation between  $\Delta_{V/2}$ and $c$ differs from
that in Alam  \etal 2002 (their eq. 9) and that  in Zentner \& Bullock
2002 (their eq.  18), both of  which are in error.  We are grateful to
Andrew Zentner for bringing this to our attention.}
\begin{equation}
\label{dvNFW}
\Delta_{V/2} = 3.36 \, \Delta_{\rm vir} \, {c^3 \over {\rm ln}(1+c) - c/(1+c)}
\end{equation}
Results  are shown  in  Figure~\ref{fig:rc}.  The  gray area  labelled
`DATA' indicates  the region occupied  by observed galaxies  (see Alam
\etal 2002).   Models \lamB, \lamC,  and \lamD predict  halo densities
inside $R_{V/2}$ that are a  factor 2, 2, and 4 smaller, respectively,
than  for  the  standard  concordance  model \lamA.   A  reduction  in
$\Omega_m$ and/or $\sigma_8$ with  respect to the standard concordance
values thus strongly alleviates  the rotation curve problem. Note that
the spread in the data is  larger than the difference between the four
cosmologies.   However,  keep  in  mind that  the  model  calculations
reflect the {\it average} $\Delta_{V/2}$ as function of $V_{\rm max}$.
Furthermore, any contribution of the baryons to $V_{\rm max}$ (ignored
here) may  boost $\Delta_{V/2}(V_{\rm max})$ of the  data with respect
to the pure dark halo models, and the curves shown here should thus be
interpreted roughly as  lower limits.  In addition, we  point out that
the  $\Delta_{V/2}(V_{\rm   max})$  shown  here  is   lower  than  the
$\Delta_{V/2}(V_{\rm  max})$ presented  in Alam  \etal (2002)  for the
same cosmology, especially at small $V_{\rm max}$.  This is due to the
different models used to  compute halo concentrations: whereas we rely
on  the  Eke \etal  (2001)  model, Alam  \etal  (2002)  use the  model
suggested  by  Bullock \etal  (2001)  which  predicts somewhat  higher
concentrations. These differences indicate the theoretical uncertainty
regarding the concentrations of  relatively low mass haloes.  Although
accurate  constraints  on cosmological  parameters  are therefore  not
possible,  the general  trend is  clear: decreasing  $\Omega_m$ and/or
$\sigma_8$ reduces the concentrations  of dark matter haloes, bringing
them in better agreement with  the data compared to the standard model
with $\Omega_m=0.3$ and $\sigma_8=0.9$.

\section{Summary}
\label{sec:concl}

One of the main goals in  modern cosmology is to determine the initial
conditions for  structure formation  in the early  Universe, expressed
through  the initial  mass power  spectrum  $P(k)$. In  this paper  we
assumed  that  $P(k)$ is  a  simple power-law  and  used  data on  the
abundance and  clustering of  galaxies to constrain  the normalization
$\sigma_8$.

Because  of the  unknown bias  of galaxies  with respect  to  the mass
distribution, previous  attempts to constrain  cosmological parameters
from large scale structure (LSS) data have mainly focussed on the {\it
shape}  of $P(k)$  rather than  the  normalization. In  this paper  we
analyzed  data  from  the  2dFGRS  using  a  technique  based  on  the
conditional luminosity  function (introduced by YMB03  and BYM03) that
self-consistently   models  the  galaxy   bias,  and   its  luminosity
dependence.   This  method,  therefore,  allows us  to  simultaneously
constrain  both the  shape and  the  normalization of  the mass  power
spectrum. In fact, unlike  additional methods to constrain $\sigma_8$,
such as cluster abundances and  weak lensing, this method allows us to
constrain both $\Omega_m$ and $\sigma_8$, rather than a combination of
both parameters.

We presented  two types of analysis.   In the first, we  focus on flat
$\Lambda$CDM cosmologies  in which only $\Omega_m$  and $\sigma_8$ are
allowed  to  vary.   The  other  parameters we  keep  fixed  at  their
``concordance''  values,  i.e., $\Omega_b  h^2  =  0,02$, $h=0.7$  and
$n_s=1.0$.  Using the luminosity function $\Phi(L)$ and the luminosity
dependence  of the  correlation lengths,  $r_0(L)$, obtained  from the
2dFGRS   by  Madgwick   \etal   (2002)  and   Norberg  \etal   (2002),
respectively, we obtain constraints  on $\Omega_m$ and $\sigma_8$ that
are in excellent agreement with  COBE.  Models with low $\Omega_m$ and
high $\sigma_8$  are robustly ruled out because  they over-predict the
amount of  clustering. Models with high $\Omega_m$  and low $\sigma_8$
are also found  to be inconsistent with the data, but  as we argued in
Section~\ref{sec:constraints}, this results mainly from our particular
parameterization of  the CLF. Adding the constraint  that the quantity
$\beta =  \Omega_m^{0.6}/b =  0.49 \pm 0.09$,  as obtained  by Hawkins
\etal (2002)  from the  redshift space distortions  in the  2dFGRS, we
obtain  that   $\Omega_m  =  0.27^{+0.16}_{-0.12}$   and  $\sigma_8  =
0.87^{+0.22}_{-0.26}$ (both 95\% CL).  These constraints, which derive
only  from  the 2dFGRS,  without  any  additional  data, are  in  good
agreement with  the constraint  $\Omega_m = 0.23  \pm 0.09$  (68\% CL)
obtained by  Hawkins \etal using  constraints on the galaxy  bias from
Verde \etal (2002)  based on an analysis of  the 2dFGRS bispectrum. It
is  reassuring that  such  wildly different  methods yield  comparable
constraints on cosmological parameters and on the bias parameter $b$.

One of  the advantages of the  CLF models presented here  is that they
allow  a  straightforward  computation  of the  average  mass-to-light
ratio,  $\langle  M_{\rm  vir}/L  \rangle_{\rm cl}$,  of  clusters  of
galaxies (defined  here as systems  with masses in excess  of $10^{14}
h^{-1} \Msun$).  We find a  very strong dependence of  $\langle M_{\rm
vir}/L \rangle_{\rm cl}$ on $\sigma_8$, which is well parameterized by
\begin{equation}
\label{sig8ml}
\langle M_{\rm vir}/L \rangle_{\rm cl} =  350 \, h \, \MLsun \, \left(
{ \sigma_8 \over 0.73} \right)^2
\end{equation}
Therefore,   any   additional,   independent   measurements   of   the
mass-to-light  ratio of  clusters of  galaxies allows  the constraints
given above to be strengthened  even further. Taking the average value
quoted in  the literature, $\langle  M_{\rm vir}/L \rangle_{\rm  cl} =
(350 \pm  70) h \MLsun$ (Carlberg  \etal 1996; Bahcall  \etal 2000) we
obtain   $\Omega_m    =   0.27^{+0.14}_{-0.10}$   and    $\sigma_8   =
0.77^{+0.10}_{-0.14}$ (both 95\% CL).  Thus the observed clustering of
galaxies,  combined with  constraints  on the  mass-to-light ratio  of
clusters, argues for a power  spectrum normalization that is $\sim 2.5
\sigma$ lower than  the standard value of $\sigma_8=0.9$.   This is in
agreement with  a rapidly growing  number of studies based  on cluster
abundances and cosmic shear measurements.  Note that our constraint on
$\sigma_8$  mainly owes to  the constraint  on $\langle  M_{\rm vir}/L
\rangle_{\rm cl}$.   Under the assumption that  $\langle M_{\rm vir}/L
\rangle_{\rm cl}$  is equal to  the universal mass-to-light  ratio, as
suggested by  the fact  that $M/L$ is  independent of scale  on scales
larger than $\sim 1 h^{-1} \Mpc$ (Bahcall \etal 1995, 2000), we obtain
that $\Omega_m  = 0.23  \pm 0.09$  (68 \% CL).   This is  in excellent
agreement  with  the  constraints  on  $\Omega_m$  given  above,  thus
indicating self-consistency.

In addition to  this analysis in which only  $\Omega_m$ and $\sigma_8$
were allowed  to vary, we also  performed a joint analysis  of the LSS
data with pre-WMAP CMB data, this time using a 6-parameter analysis of
flat  $\Lambda$CDM  cosmologies.   The  CMB data  itself  only  poorly
constrains $\sigma_8$  because of  the well-known degeneracy  with the
optical  depth due  to reionization.   Adding the  constraints  on the
galaxy  correlation   lengths  does  not   significantly  reduce  this
degeneracy. In fact,  any model that fits the CMB  data, also fits the
observed galaxy  clustering, strongly  suggesting that both  data sets
are consistent with the same matter power spectrum. However, including
$\beta = 0.49 \pm 0.09$  and $\langle M_{\rm vir}/L \rangle_{\rm cl} =
(350  \pm 70) h  \MLsun$ as  Gaussian priors  leads to  extremely well
constrained   parameters;   $\Omega_m   =  0.25^{+0.10}_{-0.07}$   and
$\sigma_8 =  0.78 \pm 0.12$ (both  95\% CL).  These  are consistent at
better  that the  $1  \sigma$ confidence  level  with the  constraints
obtained  without  the  CMB  data  using  a  more  restricted  set  of
cosmologies.  In addition, these results are in perfect agreement with
Lahav \etal (2002), who, using a similar combination of 2dFGRS and CMB
data, derived $\sigma_8 = 0.73  \pm 0.05$ (69\% CL). Clearly, the data
argues  for a relatively  low value  of $\sigma_8$,  and with  a small
preference for a matter density  of $\Omega_m \simeq 0.25$ in favor of
the concordance value of $\Omega_m  \simeq 0.3$.  This is in excellent
agreement with  recent work by  Melchiorri \etal (2003), who,  using a
combination of  CMB and  Sloan Digital Sky  Survey (SDSS)  data obtain
very similar conclusions.

Numerous  studies   have  shown  that   the  $\Lambda$CDM  concordance
cosmology  predicts   dark  matter  haloes  that   are  too  centrally
concentrated. This is apparent  from both the observed rotation curves
of dwarf and  low surface brightness galaxies and  from the zero-point
of the  Tully-Fisher relation. However, the majority  of these studies
have  focused on  cosmologies with  $\Omega_m =  0.3$ and  $\sigma_8 =
0.9$.    Lowering  $\Omega_m$  and/or   $\sigma_8$  results   in  less
concentrated dark matter haloes.  We have investigated the effect that
small  changes  in  these  two  cosmological parameters  have  on  the
aforementioned  problems.   For  a  flat $\Lambda$CDM  cosmology  with
$\Omega_m = 0.25$ and $\sigma_8 = 0.75$ (close to the values preferred
by the  analysis presented here), the halo  concentrations are reduced
by $\sim 20$  percent with respect to the  standard concordance model. 
This implies average densities inside the radius $R_{V/2}$, defined as
the radius where  the circular velocity is half  the maximum velocity,
that  are a  factor 2.5  smaller. Simple  tests show  that  this helps
significantly in solving both the rotation curve and the TF zero-point
problem.


\section*{Acknowledgements}

We  are grateful to  Antony Lewis  and Sarah  Bridle for  making their
Monte  Carlo Markov  Chains publicly  available and  to  Neta Bahcall,
Stacey  McGaugh,  Adi  Nusser,  Anna Pasquali,  John  Peacock,  Naoshi
Sugiyama,  Simon   White,  Saleem  Zaroubi  and   Andrew  Zentner  for
stimulating   discussions.  The   anonymous   referee  is   gratefully
acknowledged for  his comments that  helped to improve the  clarity of
the  paper.  FB  acknowledges  the hospitality  of  the Institute  for
Advanced Study and New York University.



\appendix

\section[]{Parameterization of the Conditional Luminosity Function}
\label{sec:AppA}

Following YMB03 and BYM03 we assume that the CLF can be described by a
Schechter function:
\begin{equation}
\label{phiLM}
\Phi(L \vert M) {\rm d}L = {\tilde{\Phi}^{*} \over \wLstar} \,
\left({L \over \wLstar}\right)^{\walpha} \,
\, {\rm exp}(-L/\wLstar) \, {\rm d}L
\end{equation}
Here   $\wLstar   =    \wLstar(M)$,   $\walpha   =   \walpha(M)$   and
$\tilde{\Phi}^{*} =  \tilde{\Phi}^{*}(M)$; i.e., the  three parameters
that describe the conditional LF depend on $M$.  In what follows we do
not explicitly write this  mass dependence, but consider it understood
that quantities with  a tilde are functions of $M$.  

We adopt  the same parameterizations  of these three parameters  as in
YMB03, which  we repeat here  for completeness. Readers  interested in
the  motivations  behind  these  particular choices  are  referred  to
YMB03. For  the total  mass-to-light ratio  of a halo  of mass  $M$ we
write
\begin{equation}
\label{MtoLmodel}
\left\langle {M \over L} \right\rangle (M) = {1 \over 2} \,
\left({M \over L}\right)_0
\left[ \left({M \over M_1}\right)^{-\gamma_1} +
\left({M \over M_1}\right)^{\gamma_2}\right],
\end{equation}
for $M  < 10^{14} h^{-1} \Msun$,  while we adopt  $\langle M/L \rangle
(M) = \langle M/L \rangle_{\rm cl}$ for $M \geq 10^{14} h^{-1} \Msun$.
This  parameterization has four  free parameters:  two normalizations,
$(M/L)_0$ and  $\langle M/L  \rangle_{\rm cl}$, a  characteristic mass
$M_1$, for  which the mass-to-light  ratio is equal to  $(M/L)_0$, and
one  slope $\gamma_1$, which  specifies the  behavior of  $\langle M/L
\rangle$  at the  low  mass end.   Note  that $\gamma_2$  is fixed  by
requiring continuity  of $\langle M/L \rangle(M)$ across  $M = 10^{14}
h^{-1} \Msun$.

A similar  parameterization is used for  the characteristic luminosity
$\wLstar$:
\begin{equation}
\label{LstarM}
{M \over \wLstar(M)} = {1 \over 2} \, \left({M \over L}\right)_0 \,
f(\walpha) \, \left[ \left({M \over M_1}\right)^{-\gamma_1} +
\left({M \over M_2}\right)^{\gamma_3}\right].
\end{equation}
Here
\begin{equation}
\label{falpha}
f(\walpha) = {\Gamma(\walpha+2) \over \Gamma(\walpha+1,1)}.
\end{equation}
with $\Gamma(x)$  the Gamma function and  $\Gamma(a,x)$ the incomplete
Gamma  function.   This   parameterization  has  two  additional  free
parameters:  a  characteristic  mass   $M_2$  and  a  power-law  slope
$\gamma_3$.  For $\walpha(M)$ we adopt:
\begin{equation}
\label{alphaM}
\walpha(M) = -1.25 + \zeta \, \log(M_{15}).
\end{equation}
Here  $M_{15}$ is  the halo  mass in  units of  $10^{15}  \Msunh$, and
$\zeta$  describes the change  of the  faint-end slope  $\walpha$ with
halo mass. 

Once   $\walpha$   and   $\wLstar$   are  given,   the   normalization
$\tilde{\Phi}^{*}$     of    the     CLF    is     obtained    through
equation~(\ref{MtoLmodel}),  using the fact  that the  total (average)
luminosity in a halo of mass $M$ is given by
\begin{equation}
\label{meanL}
\langle L \rangle(M) = \int_{0}^{\infty}  \Phi(L \vert M) \, L \, {\rm
d}L = \tilde{\Phi}^{*} \, \wLstar \, \Gamma(\walpha+2).
\end{equation}
Finally, we introduce the mass scale $M_{\rm min}$ below which the CLF
is zero;  i.e., we assume that no  stars form inside haloes  with $M <
M_{\rm min}$.  Motivated by reionization considerations (see YMB03 for
details)  we adopt  $M_{\rm min}  = 10^{9}  h^{-1}  \Msun$ throughout.

This  model for  $\Phi(L \vert  M)$ thus  contains a  total of  7 free
parameters: 2 characteristic masses; $M_1$ and $M_2$, three parameters
that describe the various mass-dependencies $\gamma_1$, $\gamma_3$ and
$\zeta$, and two normalization  for the mass-to-light ratio, $(M/L)_0$
and $\langle M/L \rangle_{\rm cl}$.
\begin{figure}
\centerline{\psfig{figure=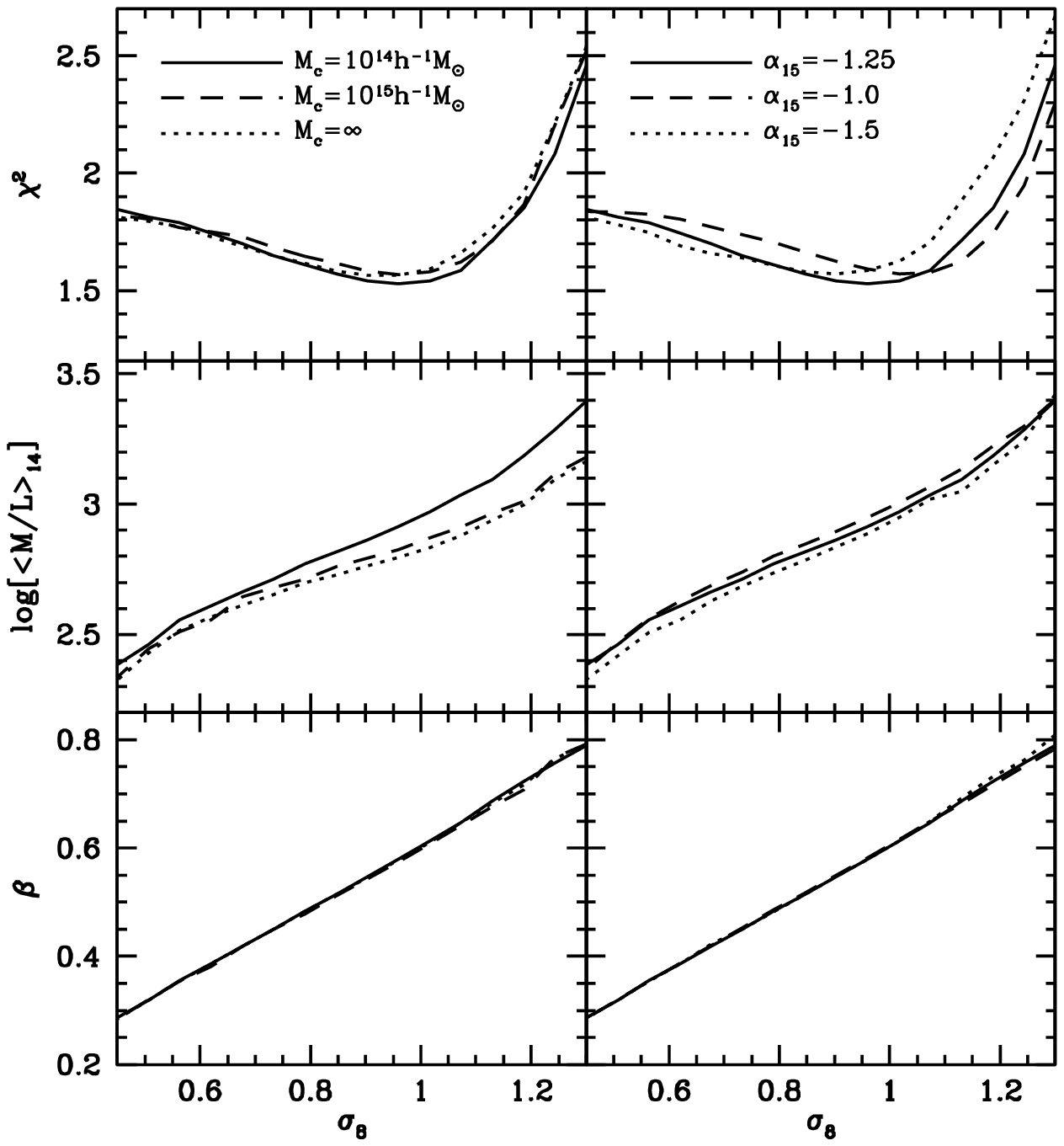,width=\hssize}}
\caption{The  dependence  of  results  on  small changes  in  the  CLF
parameterization. Panels  on the left and right  correspond to changes
in   the  parameters   $M_c$  and   $\alpha_{15}$,   respectively  (as
indicated).   Upper,  middle and  lower  panels  plot $\chi^2$,  ${\rm
log}[\langle  M/L  \rangle_{14}]$ and  $\beta$,  respectively, all  as
function  of   the  power  spectrum   normalization  $\sigma_8$.   See
Appendix~B for discussion.}
\label{fig:robust}
\end{figure}

For  the  purpose of  making  predictions  for  the TF  relation  (see
Section~\ref{sec:tulfis}), for each halo  we define a `central' galaxy
whose luminosity we denote by  $L_c$.  We assume the central galaxy to
be the brightest one in a  halo, consistent with the fact that in most
(if not all) haloes the brightest members reside near the center.  The
mean luminosity of this central galaxy is defined as
\begin{equation}
\label{Lcentral}
\langle L_c \rangle(M) = \int_{L_1}^{\infty}  \Phi(L \vert M) \, L \, 
{\rm d}L = \tilde{\Phi}^{*} \, \wLstar \, \Gamma(\walpha+2,L_1/\wLstar),
\end{equation}
with  $L_1$ defined so  that a  halo  of mass $M$   has on average one
galaxy with $L > L_1$, i.e.,
\begin{equation}
\label{Lone}
\int_{L_1}^\infty \Phi(L\vert M) dL=1\,.
\end{equation}

\section[]{Robustness of Results}
\label{sec:AppB}

One of  the  main concerns  regarding our  constraints on cosmological
parameters   is the robustness  of the  results to changes  in the CLF
model.  We have  introduced two levels  of parameterization.  First of
all, it  is  assumed  that $\Phi(L  \vert M)$   is  well fitted  by  a
Schechter form,  independent of the halo mass  $M$.  Secondly, we have
assumed various  functional forms, with a total  of 7 free parameters,
to  describe how the  three Schechter parameters ($\wLstar$, $\walpha$
and $\tilde{\Phi}^{*}$) depend on halo mass.

We  first address the robustness   of our  results against changes  in
$\wLstar$, $\walpha$,       and  $\tilde{\Phi}^{*}$   by   considering
modifications    in  our  parameterization of  $\langle   M/L \rangle$
(equation~[\ref{MtoLmodel}]).   As outlined   in   Appendix~A, in  the
fiducial model we set $\langle M/L \rangle  = \langle M/L \rangle_{\rm
cl}$ for haloes with $M \geq M_c = 10^{14} h^{-1}  \Msun$.  Thus it is
assumed that all haloes with $M \geq M_c$ have  the same {\it average}
mass-to-light  ratio.  This  is  motivated  by the fact  that  various
studies have  suggested that on cluster  mass  scales the $M/L$ varies
only weakly  with mass (e.g., Bahcall, Lubin  \&  Norman 1995; Bahcall
\etal 2000; Kochanek \etal 2002).   In order to investigate the impact
of  this assumption we compare  results for three  different values of
$M_c$: $10^{14} h^{-1} \Msun$   (the fiducial value), $M_c  =  10^{15}
h^{-1} \Msun$, and $M_c = \infty$ (i.e.,  $\langle M/L \rangle \propto
M^{\gamma_2}$ for  $M   \rightarrow   \infty$).   We  consider    flat
$\Lambda$CDM cosmologies with $\Omega_m=0.3$,  $\Omega_b h^2 =  0.02$,
$h=0.7$ and $n_s=1.0$, and compute the best-fit CLFs  for a variety of
different $\sigma_8$.    The results are shown in   the left panels of
Fig.~\ref{fig:robust}.     The       upper panel    plots     $\chi^2$
(equation~[\ref{chisq}]) as  function   of $\sigma_8$  for   all three
models (as indicated).  The middle  and lower panels plot $\langle M/L
\rangle_{14}$, the  average mass-to-light  ratio  of haloes with $M  =
10^{14} h^{-1} \Msun$, and   $\beta$, both as function of  $\sigma_8$.
Note how   $\chi^2(\sigma_8)$  and  $\beta(\sigma_8)$   are  virtually
independent of    $M_c$.  The  only quantity    that  reveals a modest
dependency on our assumption for $M_c$ is $\langle M/L \rangle_{14}$.

The panels on the right-hand side show the dependence of our models to
our parameterization   of    $\walpha(M)$.    As is    apparent   from
equation~(\ref{alphaM}), in   our fiducial  model we  set $\alpha_{15}
\equiv \walpha(10^{15} h^{-1} \Msun) = -1.25$. This again is motivated
by observations   of  the faint-end slope  of  the  LF of  clusters of
galaxies  (i.e., Beijersbergen \etal 2002;  Trentham \& Hodgkin 2002).
The    dashed and  dotted    curves    in the  right-hand  panels   of
Fig.~\ref{fig:robust} correspond   to  the best-fit  CLF   models with
$\alpha_{15}=-1.0$ and $\alpha_{15}=-1.50$, respectively. Clearly, our
choise of $\alpha_{15}$ does not have any significant impact on either
$\langle M/L \rangle_{14}$ or $\beta$. It does result in small changes
of  $\chi^2(\sigma_8)$  but   the   overal  trend remains   the  same:
cosmologies with intermediate values for $\sigma_8$ are preferred, and
cosmologies with $\sigma_8 \gta 1.2$ are clearly  ruled out. Note also
that the absolute  minimum of $\chi^2$ occurs for $\alpha_{15}=-1.25$,
which is our fiducial, observationally motivated, value.

Together with a number of  similar tests described in YMB03 and BYM03,
Fig.~\ref{fig:robust}  indicates that our  results are  robust against
(modest) changes in our  parameterization of $\wLstar$, $\walpha$, and
$\tilde{\Phi}^{*}$. This leaves, however,  the question to what extent
the assumption  of a  Schecter function for  $\Phi(L \vert  M)$ shapes
these  results. Our  motivation for  the Schechter  form  is fivefold:
first  of all,  the (conditional)  luminosity function  of  groups and
clusters (i.e.   systems with $M  \gta 10^{13} h^{-1} \Msun$)  is {\it
observed} to  be well fit by  a Schechter function  (i.e., Trentham \&
Hodgkin 2002, Muriel,  Valotto \& Lambas 1998). Second,  both the halo
mass  function, and the  (field) galaxy  luminosity function  have the
Schecter form, making it the natural functional form to choose. Third,
since the CLF  is a reflection of the  various physical processes that
play a  role during  galaxy formation, the  efficiencies of  which are
expected  to vary  smoothly with  halo  mass, it  seems reasonable  to
assume that the CLF, or  its functional form, does not change abrubtly
with halo mass. Fourth, in  YMB03 we presented an alternative form for
the CLF and showed that  this resulted in virtually identical results.
Finally, in  BYM03 we have  shown that the halo  occupation statistics
obtained  from  detailed semi-analytical  models  of galaxy  formation
compare    extremely   well    with   those    obtained    using   our
Schechter-parameterization of the CLF.  Nevertheless, neither of these
arguments    conclusively    demonstrates    that    our    particular
parameterization is appropriate  over all mass scales.  Unfortunately,
current  observational  data is  not  sufficient  to allow  completely
unparameterized  forms of  $\Phi(L  \vert M)$.   More  work, which  we
postpone  to future papers,  is therefore  required to  investigate to
what extent  alternative functional  forms for the  CLF impact  on our
results.

\label{lastpage}

\end{document}

%% file: macros_vdbosch.tex
\newcommand{\etal}{{et al.~}}

\newcommand{\kmsmpc}{\>{\rm km}\,{\rm s}^{-1}\,{\rm Mpc}^{-1}}
\newcommand{\kms}{\>{\rm km}\,{\rm s}^{-1}}

\newcommand{\Mpc}{\>{\rm Mpc}}

\newcommand{\Msun}{\>{\rm M_{\odot}}}
\newcommand{\Lsun}{\>{\rm L_{\odot}}}
\newcommand{\MLsun}{\>({\rm M}/{\rm L})_{\odot}}

\newcommand{\Msunh}{\>h^{-1}\rm M_\odot}

\newcommand{\walpha}{\tilde{\alpha}}
\newcommand{\wLstar}{\tilde{L}^{*}}

\newcommand{\lamA}{${\Lambda}30/90 \, $}
\newcommand{\lamC}{${\Lambda}25/75 \, $}
\newcommand{\lamD}{${\Lambda}20/65 \, $}
\newcommand{\lamB}{${\Lambda}30/65 \, $}

\def\gtsima{$\; \buildrel > \over \sim \;$}
\def\ltsima{$\; \buildrel < \over \sim \;$}
\def\prosima{$\; \buildrel \propto \over \sim \;$}
\def\gsim{\lower.7ex\hbox{\gtsima}}
\def\lsim{\lower.7ex\hbox{\ltsima}}
\def\simgt{\lower.7ex\hbox{\gtsima}}
\def\simlt{\lower.7ex\hbox{\ltsima}}
\def\simpr{\lower.7ex\hbox{\prosima}}
\def\la{\lsim}
\def\ga{\gsim}
\def\lta{\la}
\def\gta{\ga}

\newcommand{\apj}{ApJ}
\newcommand{\apjs}{ApJS}
\newcommand{\aj}{AJ}
\newcommand{\mnras}{MNRAS}
\newcommand{\aap}{A\&A}

\newcommand{\nat}{Nature}
\newcommand{\pasp}{PASP}

\newdimen\hssize
\newdimen\hdsize
